\documentclass[twocolumn,superscriptaddress,prd]{revtex4}
\usepackage{graphicx}
\usepackage{subfigure}
\usepackage{dcolumn}
\usepackage{bm}
\usepackage[usenames]{color}
\usepackage{array} 

\def\m{M}
\def\l{L}

\def\Y{\tilde{Y}}

\def\lsim{\mathrel{\rlap{\lower4pt\hbox{\hskip1pt$\sim$}}
   \raise1pt\hbox{$<$}}}
\def\gsim{\mathrel{\rlap{\lower4pt\hbox{\hskip1pt$\sim$}}
   \raise1pt\hbox{$>$}}}

\begin{document}
\topmargin 0.0001cm
\title{Photoproduction of $\pi^+ \pi^-$ meson pairs on the proton\\}


\newcommand*{\INFNGE}{Istituto Nazionale di Fisica Nucleare, Sezione di Genova, 16146 Genova, Italy}
\affiliation{\INFNGE}
\newcommand*{\INDIANA} {Physics Department and Nuclear Theory Center \\ Indiana University, Bloomington, Indiana 47405}
\affiliation{\INDIANA}

\newcommand*{\ANL}{Argonne National Laboratory, Argonne, Illinois 60439}
\newcommand*{\ANLindex}{1}
\affiliation{\ANL}
\newcommand*{\ASU}{Arizona State University, Tempe, Arizona 85287-1504}
\newcommand*{\ASUindex}{2}
\affiliation{\ASU}
\newcommand*{\UCLA}{University of California at Los Angeles, Los Angeles, California  90095-1547}
\newcommand*{\UCLAindex}{3}
\affiliation{\UCLA}
\newcommand*{\CSU}{California State University, Dominguez Hills, Carson, CA 90747}
\newcommand*{\CSUindex}{4}
\affiliation{\CSU}
\newcommand*{\CMU}{Carnegie Mellon University, Pittsburgh, Pennsylvania 15213}
\newcommand*{\CMUindex}{5}
\affiliation{\CMU}
\newcommand*{\CUA}{Catholic University of America, Washington, D.C. 20064}
\newcommand*{\CUAindex}{6}
\affiliation{\CUA}
\newcommand*{\SACLAY}{CEA, Centre de Saclay, Irfu/Service de Physique Nucl\'eaire, 91191 Gif-sur-Yvette, France}
\newcommand*{\SACLAYindex}{7}
\affiliation{\SACLAY}
\newcommand*{\CNU}{Christopher Newport University, Newport News, Virginia 23606}
\newcommand*{\CNUindex}{8}
\affiliation{\CNU}
\newcommand*{\UCONN}{University of Connecticut, Storrs, Connecticut 06269}
\newcommand*{\UCONNindex}{9}
\affiliation{\UCONN}
\newcommand*{\ECOSSEE}{Edinburgh University, Edinburgh EH9 3JZ, United Kingdom}
\newcommand*{\ECOSSEEindex}{10}
\affiliation{\ECOSSEE}
\newcommand*{\FU}{Fairfield University, Fairfield CT 06824}
\newcommand*{\FUindex}{11}
\affiliation{\FU}
\newcommand*{\FIU}{Florida International University, Miami, Florida 33199}
\newcommand*{\FIUindex}{12}
\affiliation{\FIU}
\newcommand*{\FSU}{Florida State University, Tallahassee, Florida 32306}
\newcommand*{\FSUindex}{13}
\affiliation{\FSU}
\newcommand*{\GWU}{The George Washington University, Washington, DC 20052}
\newcommand*{\GWUindex}{14}
\affiliation{\GWU}
\newcommand*{\ECOSSEG}{University of Glasgow, Glasgow G12 8QQ, United Kingdom}
\newcommand*{\ECOSSEGindex}{15}
\affiliation{\ECOSSEG}
\newcommand*{\ISU}{Idaho State University, Pocatello, Idaho 83209}
\newcommand*{\ISUindex}{16}
\affiliation{\ISU}
\newcommand*{\INFNFR}{INFN, Laboratori Nazionali di Frascati, 00044 Frascati, Italy}
\newcommand*{\INFNFRindex}{17}
\affiliation{\INFNFR}
\newcommand*{\INFNRO}{INFN, Sezione di Roma Tor Vergata, 00133 Rome, Italy}
\newcommand*{\INFNROindex}{19}
\affiliation{\INFNRO}
\newcommand*{\ORSAY}{Institut de Physique Nucl\'eaire ORSAY, Orsay, France}
\newcommand*{\ORSAYindex}{20}
\affiliation{\ORSAY}
\newcommand*{\ITEP}{Institute of Theoretical and Experimental Physics, Moscow, 117259, Russia}
\newcommand*{\ITEPindex}{21}
\affiliation{\ITEP}
\newcommand*{\IHEP}{Institute for High Energy Physics, Protvino, 142281, Russia}
\affiliation{\IHEP}
\newcommand*{\JMU}{James Madison University, Harrisonburg, Virginia 22807}
\newcommand*{\JMUindex}{22}
\affiliation{\JMU}
\newcommand*{\UK}{University of Kentucky, Lexington, Kentucky 40506}
\affiliation{\UK}
\newcommand*{\KHARKOV}{Kharkov Institute of Physics and Technology, Kharkov 61108, Ukraine}
\affiliation{\KHARKOV}
\newcommand*{\KYUNGPOOK}{Kyungpook National University, Daegu 702-701, Republic of Korea}
\newcommand*{\KYUNGPOOKindex}{23}
\affiliation{\KYUNGPOOK}
\newcommand*{\UMASS}{University of Massachusetts, Amherst, Massachusetts  01003}
\affiliation{\UMASS}
\newcommand*{\NINP}{Henryk Niewodniczanski Institute of Nuclear Physics PAN, 31-342 Krakow, Poland}
\affiliation{\NINP}
\newcommand*{\UNH}{University of New Hampshire, Durham, New Hampshire 03824-3568}
\newcommand*{\UNHindex}{24}
\affiliation{\UNH}
\newcommand*{\NSU}{Norfolk State University, Norfolk, Virginia 23504}
\newcommand*{\NSUindex}{25}
\affiliation{\NSU}
\newcommand*{\UNCW}{University of North Carolina, Wilmington, North Carolina 28403}
\affiliation{\UNCW}
\newcommand*{\UAT}{North Carolina Agricultural and Technical State University, Greensboro, North Carolina 27455}
\affiliation{\UAT}
\newcommand*{\OHIOU}{Ohio University, Athens, Ohio  45701}
\newcommand*{\OHIOUindex}{26}
\affiliation{\OHIOU}
\newcommand*{\ODU}{Old Dominion University, Norfolk, Virginia 23529}
\newcommand*{\ODUindex}{27}
\affiliation{\ODU}
\newcommand*{\RPI}{Rensselaer Polytechnic Institute, Troy, New York 12180-3590}
\newcommand*{\RPIindex}{28}
\affiliation{\RPI}
\newcommand*{\URICH}{University of Richmond, Richmond, Virginia 23173}
\newcommand*{\URICHindex}{29}
\affiliation{\URICH}
\newcommand*{\ROMAII}{Universita' di Roma Tor Vergata, 00133 Rome Italy}
\newcommand*{\ROMAIIindex}{30}
\affiliation{\ROMAII}
\newcommand*{\RIKEN}{The Institute of Physical and Chemical Research, RIKEN, Wako, Saitama 351-0198, Japan}
\affiliation{\RIKEN}
\newcommand*{\MOSCOW}{Skobeltsyn Nuclear Physics Institute, Skobeltsyn Nuclear Physics Institute, 119899 Moscow, Russia}
\newcommand*{\MOSCOWindex}{31}
\affiliation{\MOSCOW}
\newcommand*{\SCAROLINA}{University of South Carolina, Columbia, South Carolina 29208}
\newcommand*{\SCAROLINAindex}{32}
\affiliation{\SCAROLINA}
\newcommand*{\JLAB}{Thomas Jefferson National Accelerator Facility, Newport News, Virginia 23606}
\newcommand*{\JLABindex}{33}
\affiliation{\JLAB}
\newcommand*{\UNIONC}{Union College, Schenectady, New York 12308}
\newcommand*{\UNIONCindex}{34}
\affiliation{\UNIONC}
\newcommand*{\UTFSM}{Universidad T\'{e}cnica Federico Santa Mar\'{i}a, Casilla 110-V Valpara\'{i}so, Chile}
\newcommand*{\UTFSMindex}{35}
\affiliation{\UTFSM}
\newcommand*{\VIRGINIA}{University of Virginia, Charlottesville, Virginia 22901}
\newcommand*{\VIRGINIAindex}{36}
\affiliation{\VIRGINIA}
\newcommand*{\WM}{College of William and Mary, Williamsburg, Virginia 23187-8795}
\affiliation{\WM}
\newcommand*{\YEREVAN}{Yerevan Physics Institute, 375036 Yerevan, Armenia}
\newcommand*{\YEREVANindex}{37}
\affiliation{\YEREVAN}

\newcommand*{\NOWJLAB}{Thomas Jefferson National Accelerator Facility, Newport News, Virginia 23606}
\newcommand*{\NOWLANL}{Los Alamos National Laborotory, Los Alamos, New Mexico 87545}
\newcommand*{\NOWCNU}{Christopher Newport University, Newport News, Virginia 23606}
\newcommand*{\NOWECOSSEE}{Edinburgh University, Edinburgh EH9 3JZ, United Kingdom}
\newcommand*{\NOWWM}{College of William and Mary, Williamsburg, Virginia 23187-8795}

\author {M.~Battaglieri} 
\affiliation{\INFNGE}
\author {R.~De~Vita} 
\affiliation{\INFNGE}
\author {A.~P. Szczepaniak}
\affiliation{\INDIANA}

\author {K. P. ~Adhikari} 
\affiliation{\ODU}
\author {M.J.~Amaryan} 
\affiliation{\ODU}
\author {M.~Anghinolfi} 
\affiliation{\INFNGE}
\author {H.~Baghdasaryan} 
\affiliation{\VIRGINIA}
\author {I.~Bedlinskiy} 
\affiliation{\ITEP}
\author {M.~Bellis} 
\affiliation{\CMU}
\author {L.~Bibrzycki}
\affiliation{\NINP}
% ...

\collaboration{The CLAS Collaboration}
   \noaffiliation
%

%
%

\date{\today}

\begin{abstract}
The exclusive reaction $\gamma p \to p \pi^+ \pi^-$  was studied 
in the photon energy range 3.0 - 3.8 GeV and momentum transfer range  $0.4<-t<1.0$~GeV$^2$.
Data were collected with the 
CLAS detector at the Thomas Jefferson National Accelerator Facility. In this kinematic range
the integrated luminosity was about  20 pb$^{-1}$.
The  reaction was isolated by detecting the $\pi^+$ and proton in CLAS,
and reconstructing the  $\pi^-$ via the missing-mass technique. Moments of the di-pion decay angular distributions
were derived from the experimental data. Differential cross sections for 
the $S$, $P$, and $D$-waves in the $M_{\pi^+\pi^-}$ mass range $0.4-1.4$~GeV
were derived performing a  partial wave expansion of the extracted  moments. 
Besides the dominant  contribution of the $\rho(770)$ meson in the $P$-wave, 
evidence for the $f_0(980)$ and the $f_2(1270)$
mesons was found in the $S$ and  $D$-waves, respectively.
The differential production cross sections $d\sigma/dt$ for individual waves  in the mass range of 
the above-mentioned mesons were extracted. 
This is the first time the  $f_0(980)$ has been measured 
in a photoproduction experiment.
\end{abstract}

\maketitle

\section{\label{sec:intro}Introduction}
The two pion channel offers the possibility of investigating various aspects of the  meson resonance spectrum. 
It couples to the scalar-isoscalar channel that contains the $\sigma$, $f_0(980)$ and possibly a few 
more resonances with masses below $2\mbox{ GeV}$. 
It is the main decay mode of the  lowest isoscalar-tensor $f_2(1270)$ resonance and it is 
the only decay mode of the isovector-vector resonance, the $\rho(770)$. 
Among all these, the $\rho$-meson is by far the most prominent and most extensively studied, both from the point of view of 
its production mechanisms and its internal properties. 
Nowadays the other resonances too are subjects of  extensive theoretical and experimental investigation.
The $\sigma$ meson is now established
with  pole mass and width determined with good accuracy~\cite{Caprini:2005zr,Kaminski:2006qe,Kaminski:2006yv}.
However, its microscopic structure seems to be quite different from that of the $\rho$ and it is the subject of theoretical debate~\cite{Pelaez:2003dy}. The $f_0(980)$ is even a  more enigmatic state: its experimental determination is  complicated by its proximity to the $K{\bar K}$ threshold,
and its  QCD nature still awaits an explanation~\cite{Bugg:2004xu}. 
Finally, the $f_2(1270)$  has been represented so far as a Breit-Wigner resonance~\cite{Kaminski:2006qe} and appears to fit well into the quark model spectrum~\cite{Godfrey:1985xj}.

In this paper we focus on the scalar sector,
using the $\rho$ meson as a benchmark for the analysis procedure.
The $K{\bar K}$ channel from the same data set is currently being analyzed and in the near future   a coupled-channels analysis
will provide further constraints on the extraction of the meson properties.


For a long time
most of our knowledge on  the scalar meson spectrum was obtained from hadron-induced reactions, 
$\gamma\gamma$ collisions and studying  the  decays of various mesons,
e.g. $\phi$, $J/\Psi$, $D$ and $B$.
Very few studies were attempted with electromagnetic probes, in particular real photons,  since their production 
cross sections are relatively small compared to the dominant production 
of vector mesons. On one hand, through vector meson dominance, the photon can be effectively described as a 
virtual vector meson. On the other hand, quark-hadron duality and the point-like-nature 
of the photon coupling make it possible to describe photo-hadron interactions at the QCD level.
Recently, high-intensity and high-quality tagged-photon beams, as the one available at JLab, have opened
a new window into this field.

In photoproduction processes, information about the $S$-wave strength can be extracted 
by performing a partial wave analysis.
Angular distributions of photoproduced mesons and related observables, such as the  moments of the angular distributions
 and  the density matrix elements,  are the most effective tools
to look for interference patterns. 
An interference between the $S$-wave and the dominant $P$-wave
was discovered in  the moment analysis of $K^+K^-$ photoproduction on hydrogen, analyzing the data collected 
in the experiments performed at DESY~\cite{Behrend} 
and Daresbury~\cite{Barber}.
In two-pion production experiments, such as reported in Refs.~\cite{Ballam_1,Ballam_2,ABBHHM},
moments and density matrix elements were  used to analyze the properties of helicity amplitudes
describing the photoproduction process. Unfortunately, only the dominant spin-1 partial wave of the $\pi^+\pi^-$ pair
was taken into account. No attempt to obtain information about the $S$-wave amplitude was made.
More recently, the HERMES experiment at DESY~\cite{HERMES} investigated the interference of the $P$-wave in the $\pi^+\pi^-$
system with the $S$ and $D$-waves in the $\pi^+\pi^-$ electroproduction process, and  showed that such interference effects 
are measurable. The large photon virtuality $Q^2>$3 GeV$^2$ is, however, a crucial factor that distinguishes
this analysis from the photoproduction analysis~\cite{Ballam_1,Ballam_2}.

Theoretical models for  $\pi^+\pi^-$ photoproduction have been investigated in a series of articles. 
A very successful approach is  the one by  S\"oding~\cite{Soding} and  its numerous 
modifications~\cite{Krass,Kramer_Uretsky,Pumplin,Bauer}.
These models were able to describe  the shift
of the maximum of the $\pi^+\pi^-$ effective mass distribution with respect to the nominal $\rho$ mass and the asymmetric  shape
observed in SLAC~\cite{Ballam_1,Ballam_2} and DESY~\cite{ABBHHM,Struczinski} data.
These properties are attributed to the interference
of the dominating diffractive  $\rho$ meson production, with its subsequent decay into $\pi^+\pi^-$,
with the amplitudes corresponding to
Drell-type diagrams in which the photon dissociates into $\pi^+$ and $\pi^-$, and one of the pions 
is elastically scattered off the proton.
More recently, G\'omez Tejedor and Oset~\cite{GomezTejedor} applied an effective Lagrangian to construct the photoproduction amplitudes.
Their approach is limited to  photon energies below 800~MeV and effective masses $M_{\pi\pi}$ smaller 
than 1 GeV. A two-stage approach for the $\pi^+\pi^-$ $S$-wave photoproduction was proposed in the model 
 of Ref.~\cite{Ji}. 
First, a set of Born amplitudes, corresponding to photoproduction of  $\pi^+\pi^-$, $\pi^0\pi^0$, $K^+K^-$
and $K^0 {\bar K^0}$ pairs is calculated. Then the photoproduced meson pairs are subject to  final-state interactions resulting in
 the $\pi^+\pi^-$ system~\cite{KLM1994,KL1995,L1996,BLS}.
The coupled-channels calculations were separately performed for all isospin $I$ components of the transition matrix. 
Thus the $S$-wave amplitudes in that model account for the existence of the isoscalar $\sigma$, $f_0(980)$ and $f_0(1500)$, and the isovector
$a_0(980)$ and $a_0(1450)$ resonances. 
The coupling of the $K\bar{K}$ isovector channel with the $\pi\eta$ amplitude is described in Ref.~\cite{FL}.

All theoretical approaches described above do not consider explicitly the $s$-channel production of 
baryon resonances contributing to the  $p \pi\pi$ final state.
Data from Refs.~\cite{Ballam_1,ABBHHM,Struczinski}, as well 
as from more recent experimental studies~\cite{ELSA}, indicate that the contribution of  baryon resonances, such as 
 $\Delta^{++}$ and $\Delta^0$, dominate at 
lower incident photon  energies (below 2 GeV). Furthermore,  data obtained with
the SAPHIR detector at ELSA for  photon energies between 0.5 GeV and 2.6 GeV show that the contribution
of baryonic resonances to the $\pi^+p$ and $\pi^-p$ mass distributions gradually decreases with photon energy.

In this paper we review the results of the analysis of 
  $\pi^+\pi^-$ photoproduction in the photon energy range 3.0 - 3.8 GeV and momentum transfer 
squared $-t$  between 0.4~GeV$^2$ and 1~GeV$^2$, where
the di-pion effective mass $M_{\pi\pi}$ varies from 0.4 GeV to  1.4 GeV. 
The main results were previously reported in Ref.~\cite{PRL_f0}. 
We are not aware of any previous evidence of  scalar mesons, in particular of the $f_0(980)$, in photoproduction of pion pairs.
This effective mass region is dominated by the production of the $\rho(770)$ resonance in the $P$-wave. 
From other experiments, such as pion-nucleon collisions $\pi^-p\to\pi^+\pi^-n$~\cite{Grayer,Becker} 
or nucleon-antinucleon annihilation~\cite{Amsler1}, there is some evidence that  resonant states  are formed in  the $S$-wave. 
These resonances have been neglected in previous experimental analyses of $\pi^+\pi^-$ photoproduction and, to our knowledge,
the current analysis is the first one that explicitly takes into account the possibility that the $S$-wave is produced 
in the $\pi^+\pi^-$ system. 

In the following, some details are given on the experiment and data analysis (Sec.~\ref{sec:exp}),
on the extraction of the angular moments of the di-pion system (Sec.~\ref{sec:mom}), and the 
fit of the moments using a dispersion relation (Sec.~\ref{sec:disp}).
Results of the partial wave analysis (differential cross section for each partial wave and 
the spin density matrix elements) and the physics interpretation are reported in Sec.~\ref{sec:res}.

\section{\label{sec:exp} Experimental procedures and data analysis}
\subsection{The photon beam and the target}
The measurement   was performed using the CLAS detector~\cite{B00} 
in  Hall B at Jefferson Lab with a bremsstrahlung 
photon beam  produced by a continuous 60-nA electron beam of energy  $E_0$ = 4.02 GeV  
impinging on a gold foil of thickness $8 \times 10^{-5}$ radiation lengths.
A bremsstrahlung tagging system~\cite{SO99} with a photon energy resolution of 0.1$\%$ $E_0$ 
was used to tag photons in the energy range from 1.6 GeV
to a maximum energy of 3.8 GeV. In this analysis only the high-energy 
part of the photon spectrum, ranging from 3.0 to 3.8 GeV, was used.
$e^+$ $e^-$ pairs produced by the interaction of the  photon beam on a thin gold foil were used
to continuously monitor the photon flux  during the experiment. Absolute normalization was obtained by comparing 
the $e^+$ $e^-$ pair rate with the photon flux measured by a total absorption lead-glass counter in dedicated
low-intensity runs.
The  energy calibration of the Hall-B tagger system
was performed both by a direct measurement of the $e^+e^-$ pairs produced by the incoming photons~\cite{tag-abs_cal}
and by applying  an over-constrained kinematic fit  to the  reaction $\gamma p \to p \pi^+ \pi^-$, where all particles
in the final state were detected in CLAS~\cite{tag-kinefit}.
The quality of the calibrations was checked by 
looking at  the mass  of known particles, as well as their dependence on  other kinematic variables 
(photon energy, detected particle momenta and angles).

The target cell, a Mylar  cylinder  4 cm in diameter and 40-cm long, was filled by liquid hydrogen at 20.4 K.
The luminosity was obtained as the product of the target density, 
target length and the incoming photon flux 
corrected for data-acquisition dead time.
The overall   systematic uncertainty on the run  luminosity was estimated to be in the range of  10$\%$,
dominated by the uncertainties on the photon flux.

\begin{figure}[h]
\vspace{8.5cm}
\includegraphics{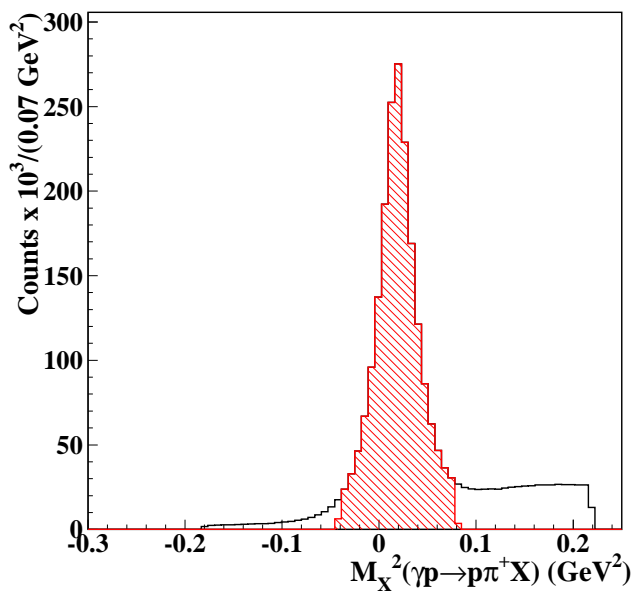}
\caption[]{Missing mass squared for the reaction $\gamma p \to p \pi^+ X$ and the $\pi^-$ peak. 
The shaded area indicates the retained events.}
\label{fig:pid}
\end{figure}

\subsection{The CLAS detector}
Outgoing  hadrons were detected in the CLAS  spectrometer.
Momentum information for charged particles was obtained via tracking
through three regions of multi-wire drift chambers~\cite{DC} within  a toroidal magnetic 
field ($\sim 0.5$ T) generated by six superconducting coils. 
The polarity of the field was set to bend the positive particles away from the beam line  into the acceptance 
of the  detector.
Time-of-flight scintillators (TOF) were used for charged hadron
identification~\cite{Sm99}. 
The interaction time between the incoming photon and the target
was measured by  the start counter (ST)~\cite{ST}. This   is
made of 24 strips of 2.2-mm thick plastic scintillator surrounding the hydrogen cell
with a single-ended PMT-based read-out. 
A time resolution of $\sim$300 ps  was achieved.

The CLAS momentum resolution, $\sigma_p/p$, ranges from 0.5 to 1\%, depending on
the kinematics. 
The detector geometrical acceptance for each positive particle in the 
relevant kinematic region is about 40\%. It is somewhat less for low-energy negative 
hadrons, which can be lost at  forward angles because
their paths are bent toward the beam line and out of the acceptance
by the toroidal field.
Coincidences between the photon tagger and the CLAS detector triggered 
the recording of the events. The trigger in CLAS  required
a coincidence between the TOF and the ST 
in at least two sectors, in order to  select
reactions with at least two charged particles in the final state.
An integrated luminosity of 70 pb$^{-1}$ ($\sim20$ pb$^{-1}$ in the range 3.0$<E_\gamma<$3.8 GeV)
was accumulated in  50 days of running  in 2004. 

\begin{figure}[h]
\vspace{8.5cm} 
\includegraphics{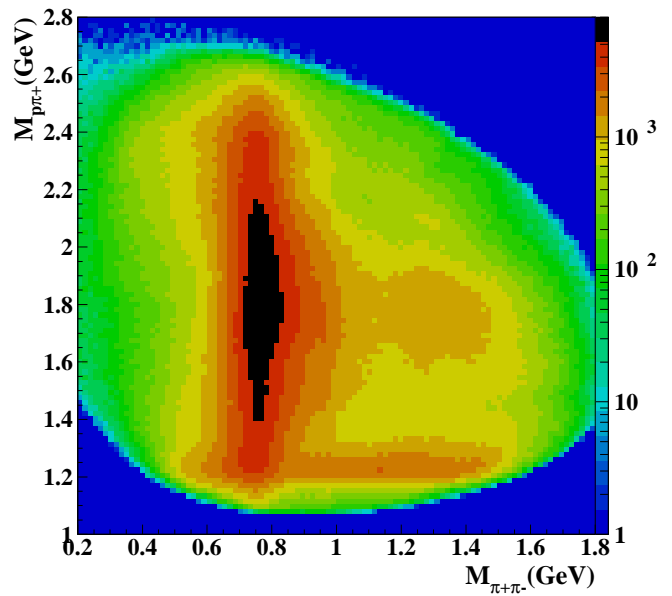}
\caption[]{Two dimensional plot of the invariant masses obtained combining pairs of  particles of the exclusive reaction
 $\gamma p \to p \pi^+ \pi^-$.} 
\label{fig:dalitz}
\end{figure}

\begin{figure}[h]
\vspace{8.5cm} 
\includegraphics{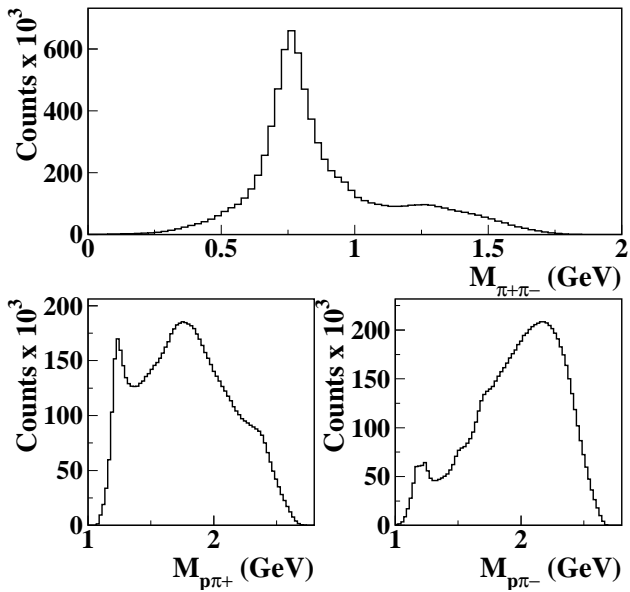}
\caption[]{Invariant masses obtained combining pairs of  particles of the exclusive reaction
 $\gamma p \to p \pi^+ \pi^-$. Upper panel $M_{\pi^+\pi^-}$; lower panel left  $M_{p\pi^+}$;   lower panel right  $M_{p\pi^-}$.
Spectra are not corrected for the detector acceptance.} 
\label{fig:invmasses}
\end{figure}

\subsection{Data analysis and reaction identification}\label{ssec:reac_id}
The raw data were passed through the standard CLAS reconstruction software to determine the four-momenta of detected particles.
In this phase of the analysis, corrections were applied to account for the energy loss of charged particles in the target and 
surrounding  materials, misalignments of the  drift chamber's positions, and 
uncertainties in the value of the toroidal magnetic field.

The reaction $\gamma p \to p \pi^+ \pi^-$ was isolated detecting the  proton and the $\pi^+$ in the CLAS spectrometer,
while the  $\pi^-$ was reconstructed from the four-momenta of the detected particles by using the missing-mass technique.
In this way the exclusivity of the reaction is ensured,  
keeping the contamination from the  multi-pion background to a minimum. Figure~\ref{fig:pid}
shows the $\pi^-$ missing mass squared.
The  background below the missing pion peak appears as a smooth contribution 
in the $\pi \pi$ invariant mass without creating narrow structures.

To avoid  edge regions in the  detector acceptance, only events within a {\it fiducial} volume were retained in this analysis.
In the laboratory reference system, cuts were defined for
the minimum hadron momentum ($p_{proton}>0.32$~GeV and $p_{\pi^+}>0.125$~GeV), and the minimum  and maximum
azimuthal angles ($\theta_{proton,\pi^+} >10^\circ$ and $\theta_{\pi^+} <120^\circ$). 
The  {\it fiducial} cuts  were defined comparing in detail the experimental data distributions with the results of  the detector  simulation.
The minimum momentum cuts were tuned for different hadrons to take into account the energy loss by ionization of the particles. 

After all cuts, 41M  events were identified as produced in the exclusive reaction   $\gamma p \to p \pi^+ \pi^-$.
The other event topologies, with at least two hadrons in the final state ($p \pi^-$, $\pi^+\pi^-$,
$p \pi^+  \pi^-$), were not used since in the kinematics of interest for this analysis ($-t<1$~GeV$^2$),
the collected data are  about one  order of magnitude less due to the detector acceptance.
Figures ~\ref{fig:dalitz} and ~\ref{fig:invmasses} show the invariant mass spectra of the different combinations of particles in the final state.
The $\rho(770)$ dominates the $\pi \pi$ spectrum and  the $\Delta(1232)^{++}$ peak is clearly visible in the $p \pi^+$
invariant mass. Figure 2 shows  a small overlap between the $\Delta(1232)^{++}$ and the $\pi \pi$ spectrum.
Baryonic resonances in the $p \pi^-$ invariant mass spectrum are less pronounced. 
It has to be noted that the projection of the baryon resonance  peaks in the  $\pi \pi$ spectrum results in a smooth contribution
and cannot create  narrow structures. The effect of this background was extensively studied as discussed in Sec.\ref{sec:sys}.

\section{\label{sec:mom} Moments of the di-pion angular distribution}\label{par:fin_results} 
In this section we consider the analysis of moments of the  di-pion angular distribution defined as:
\begin{equation}\label{eq:mom}
\langle Y_{\l\m} \rangle(E_\gamma,t,M_{\pi\pi}) = \sqrt{4\pi} \int d\Omega_\pi  {{d\sigma} \over {dt dM_{\pi\pi} d\Omega_\pi}} Y_{\l\m}(\Omega_\pi),
\end{equation} 
where $d\sigma$ is the differential cross section (in momentum transfer $t$ and di-pion invariant mass $M_{\pi\pi}$), $Y_{\l\m}$ are
spherical harmonic functions of degree $\l$ and order $\m$,  and 
$\Omega_\pi = (\theta_\pi , \phi_\pi)$ are the polar and azimuthal angles of the $\pi^+$ flight direction
in the $\pi^+\pi^-$ helicity rest frame. For the definition of the angles in the di-pion system we follow the convention of Ref.~\cite{Ballam_1}.
It follows from  Eq.~\ref{eq:mom} that,
for a given $E_\gamma,t$ and di-pion mass $M_{\pi\pi}$,
$\langle Y_{00}\rangle$ corresponds to the di-pion production  differential 
cross section $d\sigma/dtdM_{\pi\pi}$. 

There are many advantages in defining and analyzing moments rather than proceeding via a direct partial wave fit of the angular distributions.
Moments  can  be  expressed as bi-linear in terms of the partial waves and, 
depending on the particular combination of $L$ and $M$,  show specific sensitivity to  a particular subset of them.
In addition, they can be directly and
unambiguously derived from the data, allowing for a quantitative comparison to the same observables calculated in specific theoretical models.

Extraction of moments requires that the measured angular distribution is corrected by the detector acceptance.
We studied three methods for implementing acceptance corrections. 
In the first two  methods, the moments were expanded in a model-independent way in a set of basis functions and,
after weighting with Monte Carlo events,
they were compared to the data by  maximizing a likelihood function.
The first of these two parametrizes the theory in terms of 
simplified $amplitudes$, while the second uses directly $moments$ as defined above.
The approximations in these methods have to do with the choice of the basis and depend on the number of basis functions used.
The systematic effect of such truncations was studied and the main results are reported below. 
In the last method,  data and Monte Carlo were binned
in all kinematical variables. The data were then corrected by the acceptance defined as the ratio of 
reconstructed over  generated Monte Carlo events in that bin.
Since it was found  to be not reliable in bins  where the acceptance was small or vanishing, this method  was only used as a
check of the others and was
not included in the 
final determination of the experimental moments.

\subsection{Detector efficiency}
The CLAS detection efficiency for the reaction  $\gamma p \to p \pi^+\pi^-$ was obtained by means of detailed Monte Carlo 
studies, which included knowledge of the full detector geometry and a realistic response to traversing particles. Events were generated 
according  to three-particle phase space with a  bremsstrahlung photon energy spectrum.
A total  of $4$ billion  events were generated in the energy range  3.0 GeV $< E_\gamma <$  3.8 GeV  and covered
the allowed kinematic  range in $-t$ and $M_{\pi \pi}$. About 700M  events were reconstructed 
in the $M_{\pi \pi}$ and $-t$ ranges of interest
(0.4 GeV $< M_{\pi \pi}<$ 1.4 GeV , 0.1 GeV$^2< -t < $ 1.0 GeV$^2$). 
This corresponds to more than fifteen times the  statistics collected in the experiment, thereby introducing a negligible 
statistical uncertainty with respect to the statistical  uncertainty of the data.

\subsection{Extraction of the moments via likelihood fit of experimental data}
Moments were derived from the data using detector efficiency-corrected fitting functions.
As mentioned above, the expected theoretical  yield was parametrized in terms of appropriate physics functions: production amplitudes
in one case and  moments of the cross section in the other. The theoretical expectation, after correction  for acceptance, 
was  compared to the experimental yield. 
Parameters were extracted by maximizing a likelihood function defined as:
\begin{equation}
{\cal L} \sim   \Pi_{a=1}^n  \left[ {\eta(\tau_a){I(\tau_a) } \over {\int d\tau \eta(\tau) I(\tau) }} \right]. 
\end{equation} 
Here  $a$ represents a data event, $n = \Delta N$ is the number of data events in a given $(E_\gamma,-t,M_{\pi \pi})$ bin ({\it i.e.}
the fit is done independently in each bin), $\tau_a$ represents  the set of kinematical variables of the $a^{th}$ event,
 $\eta(\tau_a)$ is the corresponding acceptance derived by Monte Carlo simulations  and $I(\tau_a)$ is the theoretical 
function representing the expected event distribution.
The measure $d\tau$ includes the phase space factor and the likelihood function is normalized to the expected number of events in the bin
\begin{equation} 
{\bar n} = \int d\tau \eta(\tau) I(\tau). 
\end{equation} 
The advantage of this approach lies in avoiding binning the data and the large uncertainties related to the corrections
in  regions of CLAS with vanishing efficiencies. Comparison of the results of the two different parametrizations 
allows one to estimate the systematic uncertainty related to the procedure.
In the following, we describe the two approaches in more detail.

\subsubsection{Parametrization with amplitudes}
The expected theoretical yield in each  bin is described as:
\begin{eqnarray} 
I(\tau_a) = 4\pi \left| \sum_{\l=0}^{\l_{max}}\sum_{M=-\l}^{\l} 
a_{\l\m}(E_\gamma,-t,M_{\pi\pi})  Y_{\l\m}(\Omega_\pi)  \right|^2.  \label{partial} 
\end{eqnarray}
This parametrization has the benefit  that the intensity function $I(\tau_a)$ is by construction positive.
However, it can lead to ambiguous results since it has more parameters than  can be determined from the data.
In addition, for practical reasons, the parametrization involves a cutoff, $\l_{max}$, in the maximum number of amplitudes.
For a specific choice of $\l_{max}$, the number of fit parameters is given by $2(\l_{max}+1)^2$.
We also note that these amplitudes are not the same as the partial wave amplitudes in the usual  sense of a di-pion 
photoproduction amplitude, since  the latter depends on the nucleon and photon spins. 

After removing the irrelevant constants, the fit is performed minimizing the function:
\begin{eqnarray}
- \ln {\cal L} &=& - \sum_{a=1}^{\Delta N} \ln \eta(\tau_a) I(\tau_a) \\ \nonumber
&+&  \Delta N  \ln \sum_{\l'\m';\l\m} 
{\tilde a}^*_{\l'\m'} {\tilde a}_{\l\m} \Psi_{\l'\m';\l,\m},
\end{eqnarray}
where we have  introduced the rescaled amplitudes  ${\tilde a}_{\l\m}(E_\gamma,-t,M_{\pi \pi})$ defined by: 
\begin{equation}
{\tilde a}_{\l\m} = \sqrt{   {\eta}  } a_{\l\m},
\end{equation} 
and the acceptance matrix $\Psi(E_\gamma,-t,M_{\pi \pi})$ was computed using Monte Carlo events as:
\begin{equation}
\eta \Psi_{\l'\m';\l\m} =  {{4\pi } \over {\Delta N_{Gen} }} \sum_{a=1}^{\Delta N_{Rec}} 
Y^*_{\l'\m'}(\Omega_\pi) Y_{\l\m}(\Omega_\pi),
\end{equation} 
where $\Delta N_{Gen}$ and $\Delta N_{Rec}$ are the number of generated and reconstructed events, respectively.

Fits were done using MINUIT with the analytical expression for the gradient, and using the SIMPLEX procedure followed by MIGRAD~\cite{minuit}. 
After each fit, the covariance matrix was checked and if it was not positive definite, the fit was restarted with random input parameters.
At the end, the uncertainties were computed from the full covariance matrix.

\subsubsection{Parametrization with moments}
The  expected theoretical  yield in each $(E_\gamma,-t,M_{\pi \pi})$ bin is described as:
\begin{eqnarray} 
 I(\tau_a) =  \sqrt{4\pi}  \sum_{\l=0}^{\l_{max}}\sum_{M=0}^{\l}  \langle \Y_{\l\m}   \rangle \mbox{ Re}Y_{\l\m}(\Omega_\pi).
\end{eqnarray}
The parametrization in terms of the moments directly gives the quantities we are interested in (moments $\langle \Y_{\l\m} \rangle$).
However, the fit has to be restricted to make sure the intensity is positive.
As in the amplitude parametrization,  a cutoff $\l_{max}$ in the maximum number of moments has to be used.
The number of fit parameters is given by $(\l_{max}+1)(\l_{max}+2)/2$.
As $\l_{max}$  increases, moments with $L$ close to $\l_{max}$ show a significant variation,
while  moments with the lowest $L$ remain unchanged.


The expected (acceptance-corrected) distribution is  then given by:
\begin{equation} 
I(\tau_a)  =  \sqrt{4\pi} \sum_{\l,\m}  \left[ \eta_{\l\m} \mbox{ Re} Y_{\l\m}(\Omega_\pi) \right]  \langle \Y_{\l\m} \rangle.
\end{equation} 
The function to be minimized with respect to $\langle \Y_{\l\m} \rangle$ ($\l > 0$) 
is then given by:
\begin{equation} 
- 2\ln {\cal L} =  -2 \sum_{a=1}^{\Delta N} \ln I(\tau_a), 
\end{equation} 
with the  coefficients  $\eta_{\l\m}(E_\gamma,-t,M_{\pi \pi})$ computed using Monte Carlo  events 
\begin{equation} 
\eta_{\l\m}= {{\sqrt{4\pi} }\over {\Delta N_{Gen}}} \sum_i^{\Delta N_{Rec} } {{\mbox{ Re} Y_{\l\m}(\Omega_i)  } \over {\epsilon_\l}},
\end{equation} 
where $\epsilon_\l = 1$ for $\l=0$ and $1/2$ for all other $(\l\m)$.
For $\l_{max} \le 4$, the results are similar to what was obtained with the previous method, showing the 
same stability against  $\l_{max}$ truncation and a similar
goodness of the fit.
To check the sensitivity of the likelihood fit to the parameter initialization, moments were extracted in three different ways:
1) using random initialization for all parameters; 
2) fixing the parameters up to  $\l=2$ to the ones obtained from a fit with $\l_{max}=2$, and  randomly initializing the others;
3) starting with parameters obtained in 2) and then releasing all parameters.
The three
different methods gave consistent results and the difference of moments obtained  using the different procedures was used to evaluate 
the systematic uncertainty related to the fit procedure.

\begin{figure}
\vspace{13.cm}
\includegraphics{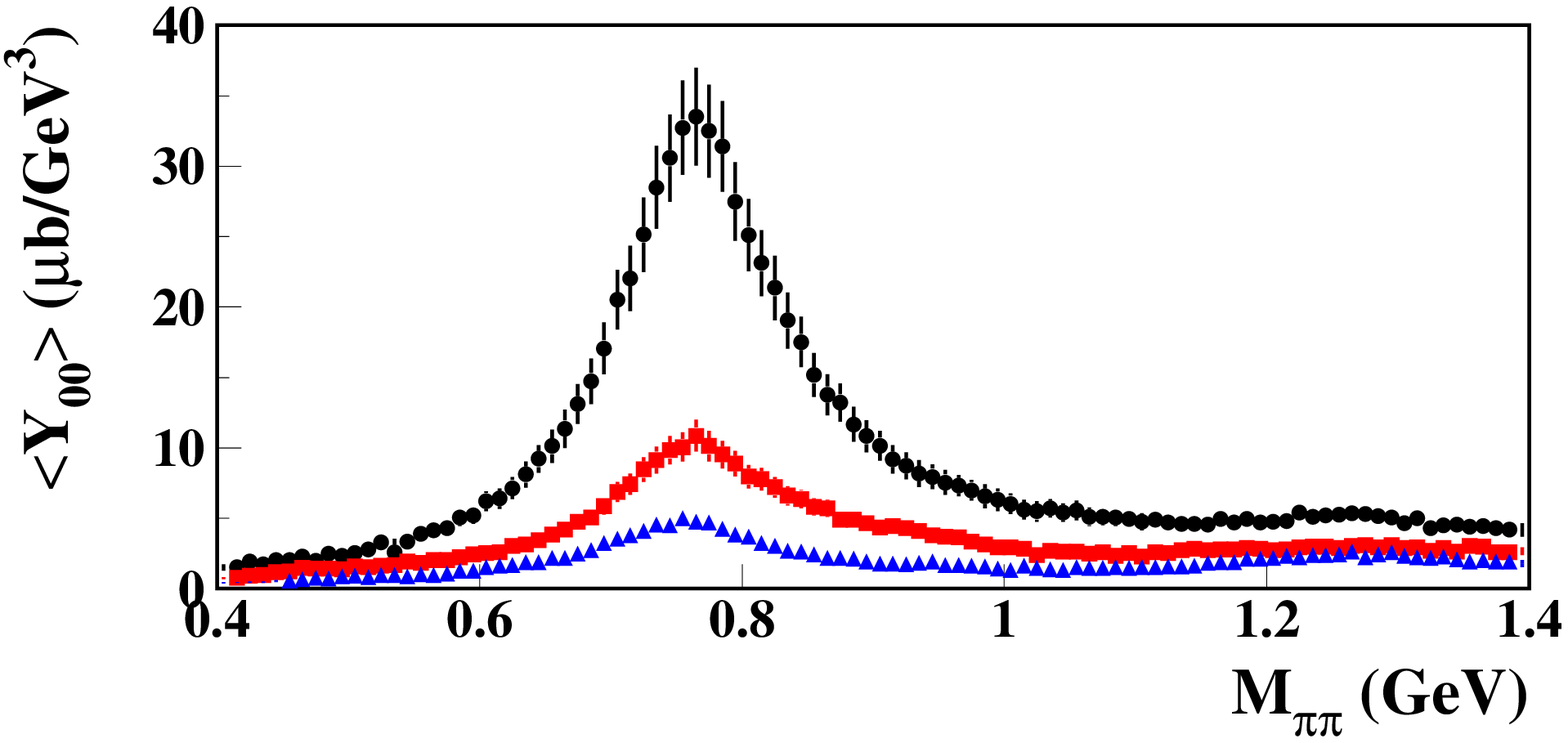}
\includegraphics{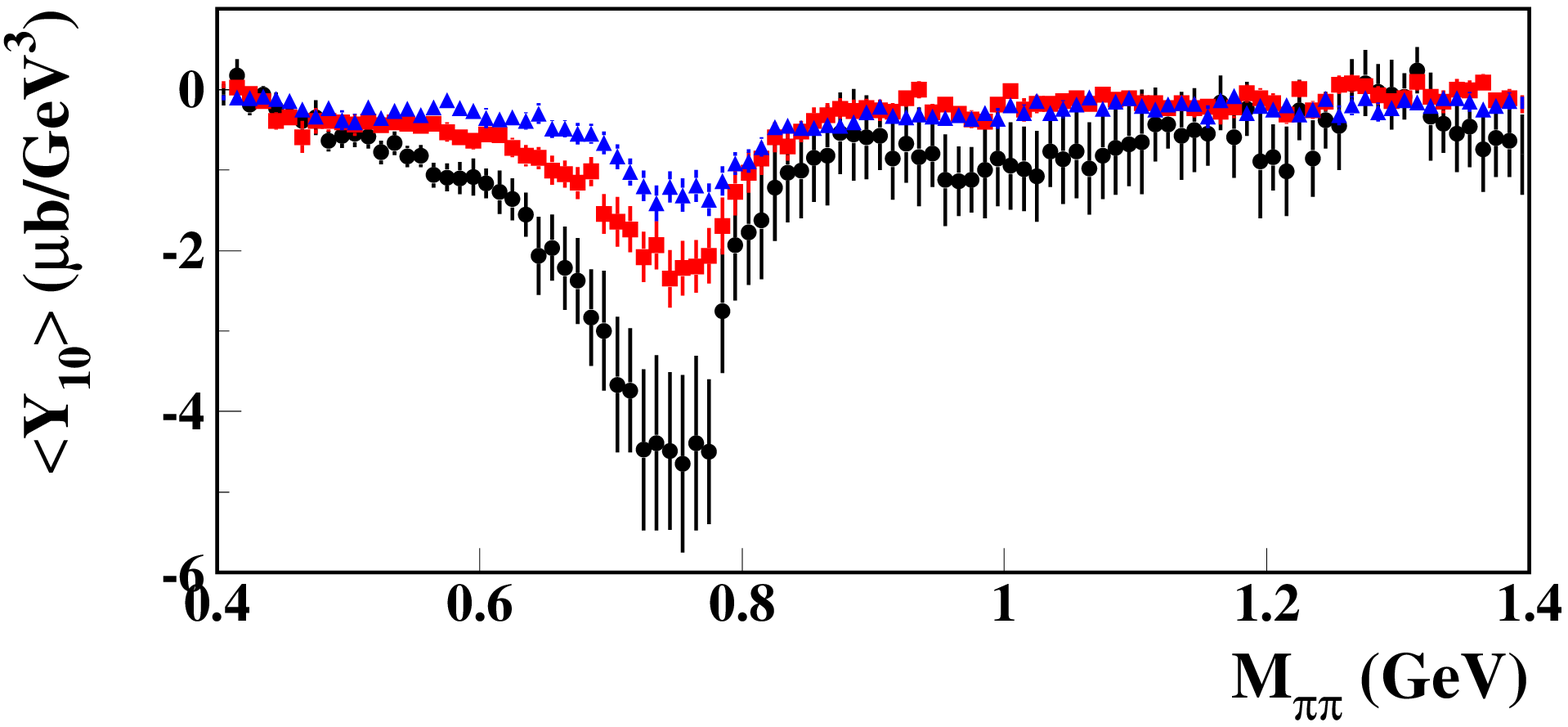}
\includegraphics{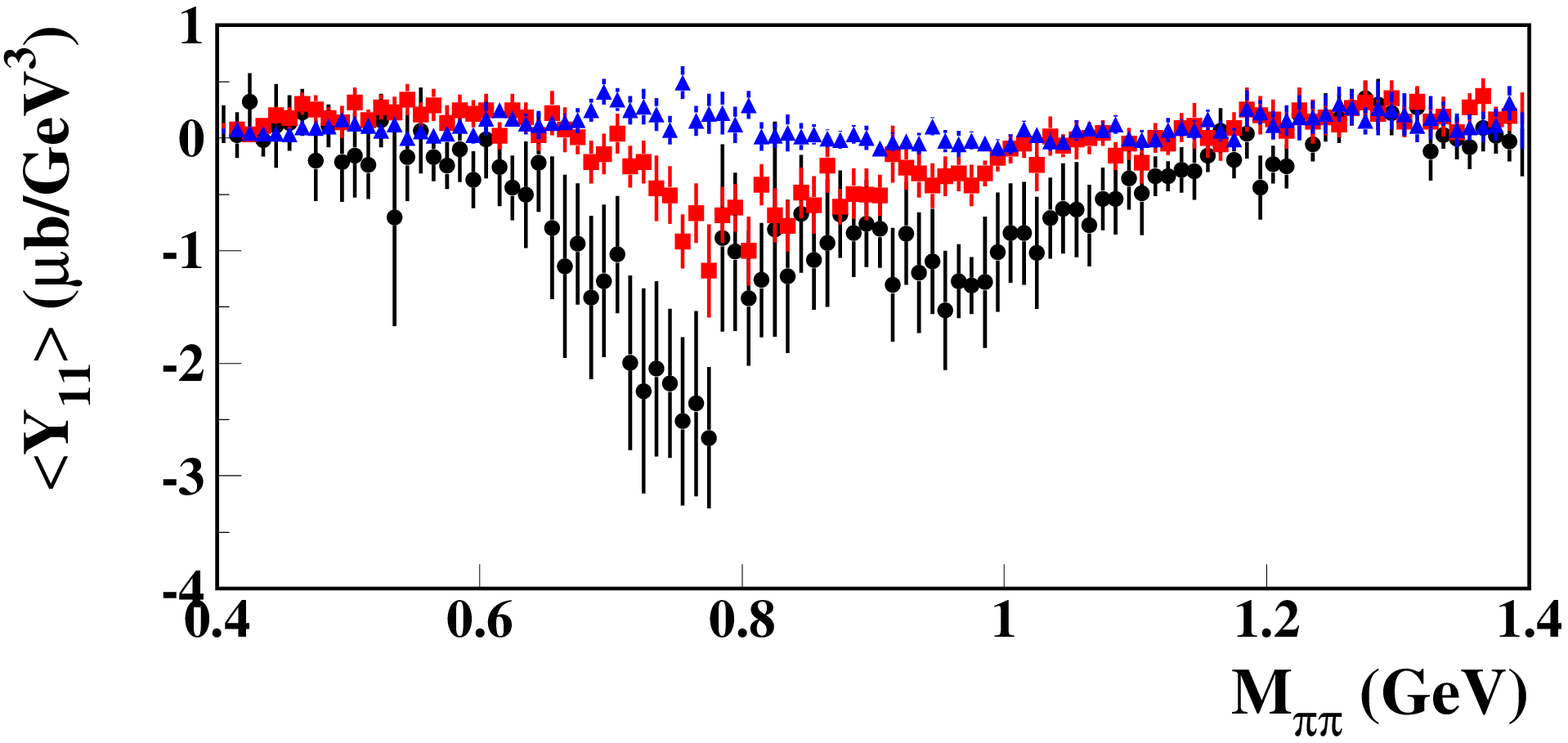}
\caption[]{Moments of the di-pion angular distribution  in $3.2 <E_\gamma< 3.4$~GeV and $-t=0.45\pm0.05$~GeV$^2$ (black), 
$-t=0.65\pm0.05$~GeV$^2$ (red) and $-t=0.95\pm0.05$~GeV$^2$ (blue). Error bars include both statistical and systematic uncertainties as explained in the text. }
\label{fig:final-1}
\end{figure}

\subsubsection{Methods comparison and final results}\label{par:syserrmom}
Moments derived by the different procedures agreed qualitatively.
The most stable results were obtained by using the first  parametrization,
although we do find occasionally large bin-to-bin fluctuations.
However, there are no a priori reasons to prefer one of the two methods and 
we consider the discrepancies between the fit results as a good estimate of the systematic uncertainty associated with
the moments extraction. 
The final results are  given as the  average of the
first method (parametrization with amplitudes) and the second  method (parametrization with moments) with the three fit initializations:
\begin{equation}
Y_{final}={{1}\over{4}}\sum_{i=1,4 \, Methods}{Y_i},
\end{equation}
where $Y$ stands for  $\langle Y_{\l\m} \rangle(E_\gamma,t,M_{\pi\pi})$.

The total uncertainty on the final moments was evaluated adding in quadrature
the statistical uncertainty, $\delta Y_{MINUIT}$ as given by MINUIT, and two systematic uncertainty contributions:
$\delta Y_{syst\,\, fit}$ related to the moment extraction procedure, 
and $\delta Y_{syst\,\,  norm}$, the systematic uncertainty  associated with the photon flux normalization (see Sec.~\ref{sec:exp}).
\begin{eqnarray}
\delta Y_{final}=\sqrt{\delta Y_{MINUIT}^2+\delta Y_{syst\,\, fit}^2+\delta Y_{syst\,\, norm}^2}\label{eq:err_Y}
\end{eqnarray}
with:\\
\begin{eqnarray}
\delta Y_{syst\,\,  fit}&=&\sqrt{\sum_{i=1,4 \, Methods}{({Y_i}-Y_{final})^2\over{4-1}}}\\
\delta Y_{syst\,\,  norm}& =& 10\% \cdot Y_{final}.
\end{eqnarray}

\begin{figure}
\vspace{13.cm}
\includegraphics{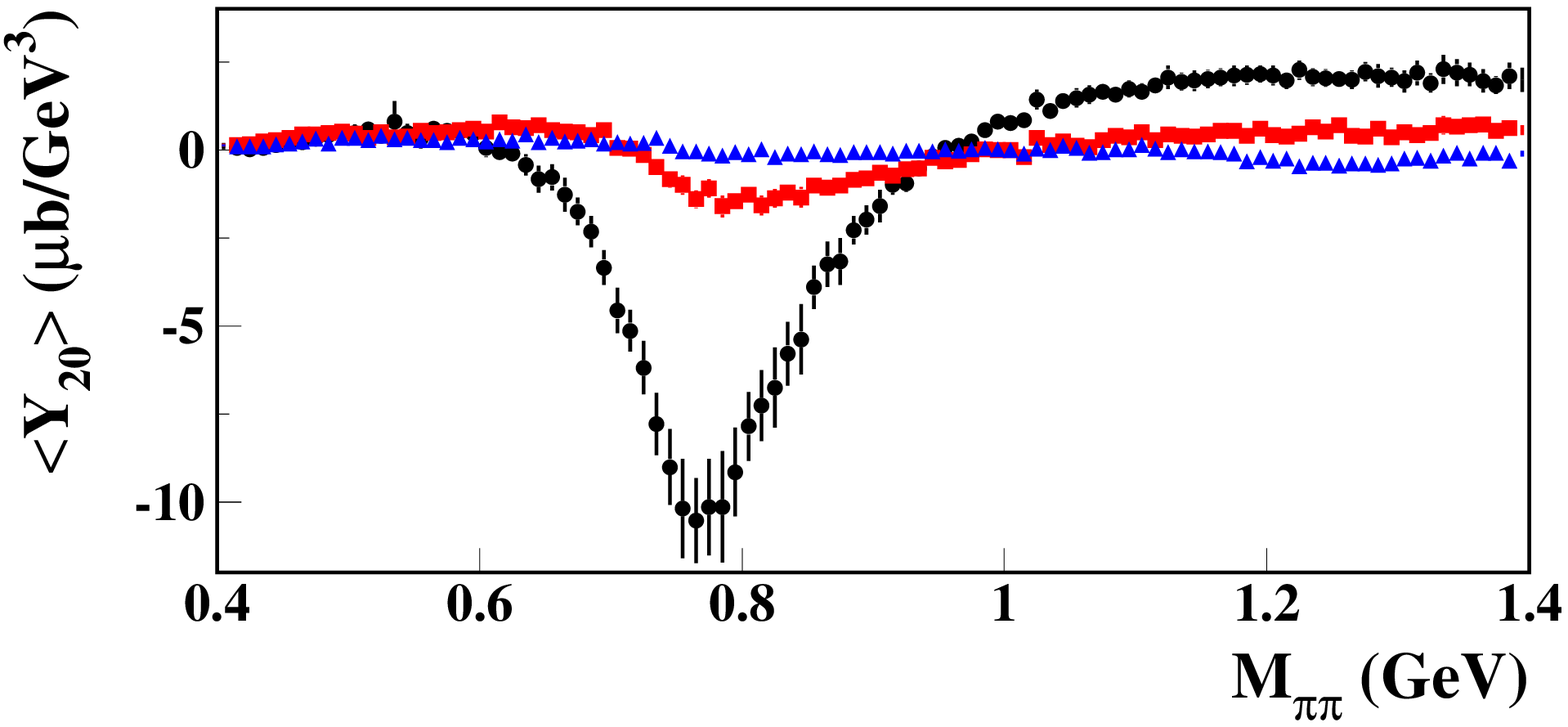}
\includegraphics{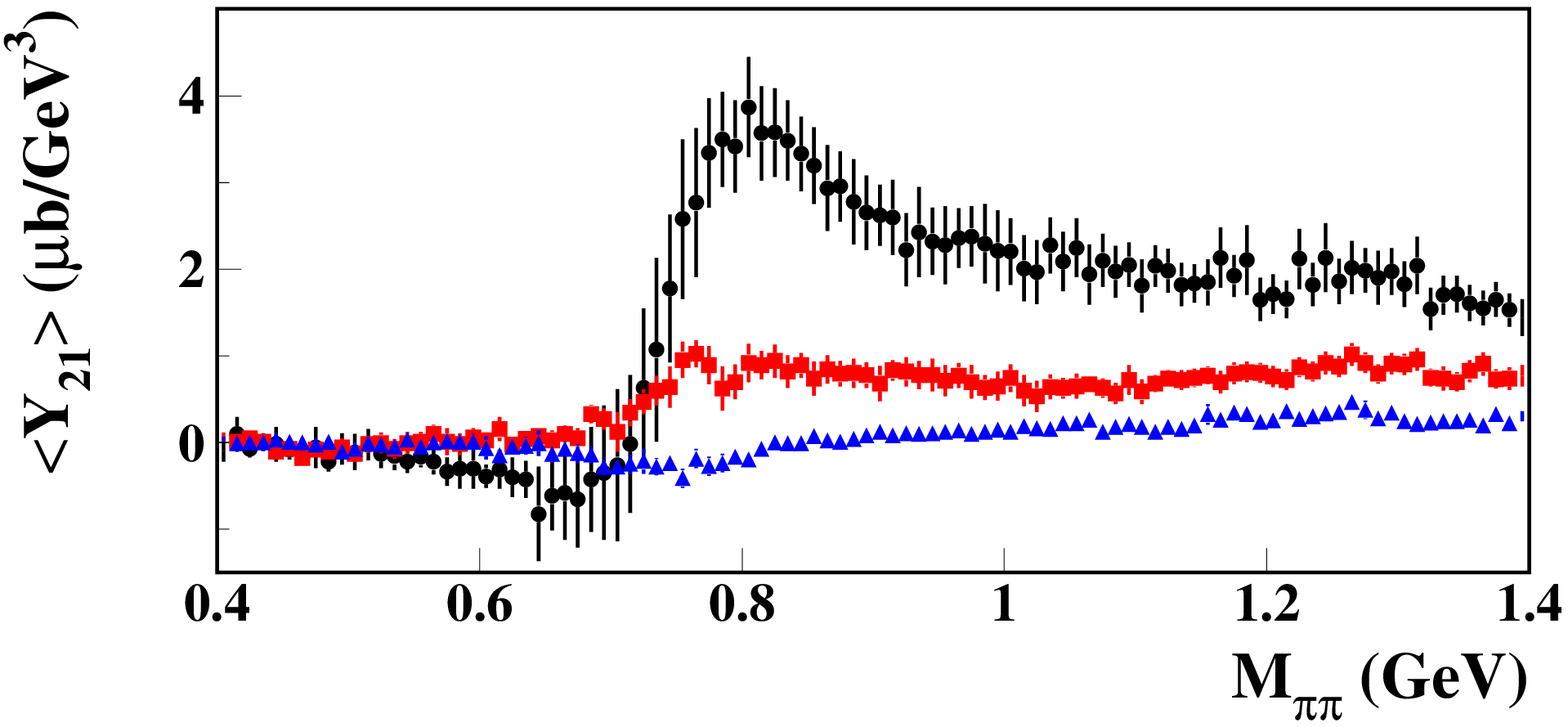}
\includegraphics{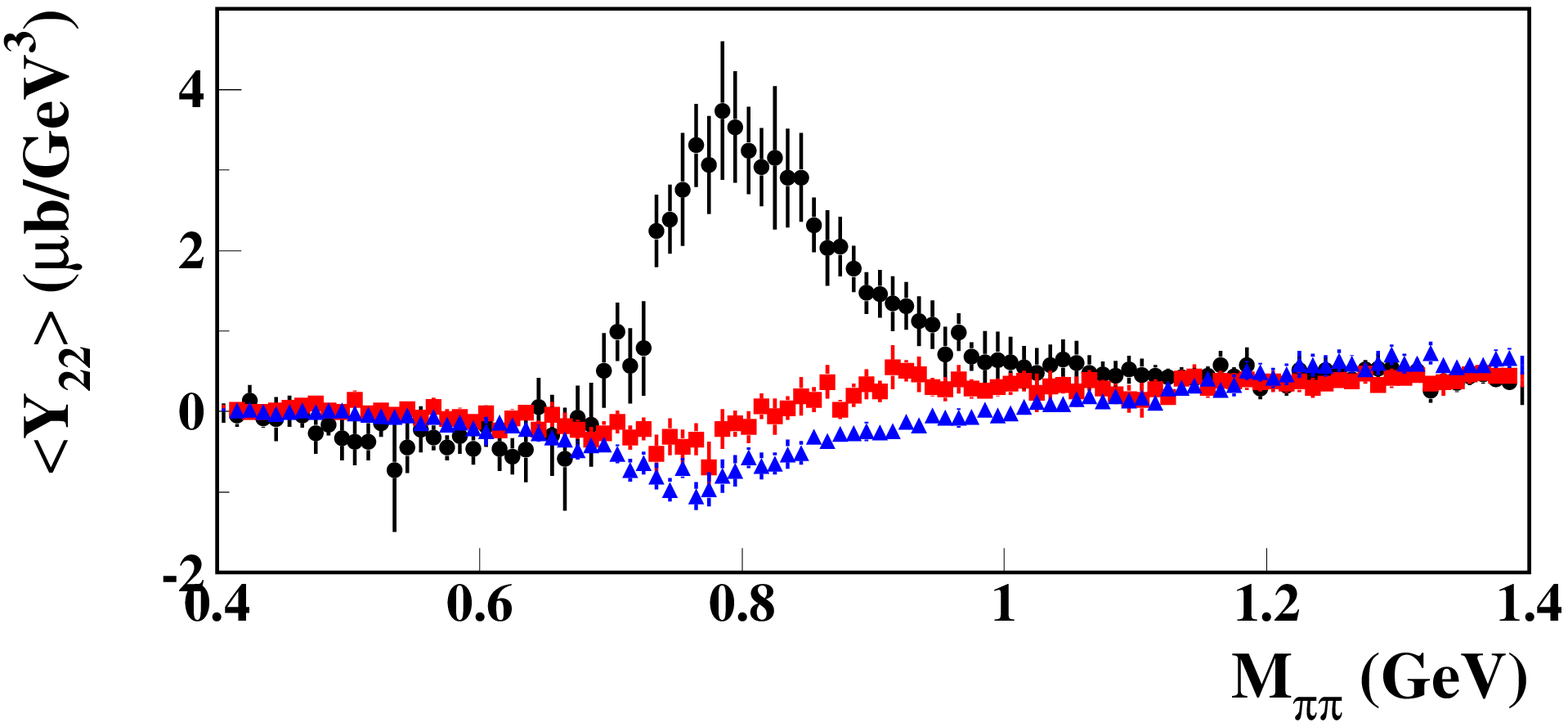}
\caption[]{Moments of the di-pion angular distribution  in $3.2 <E_\gamma< 3.4$~GeV 
and $-t=0.45\pm0.05$~GeV$^2$ (black), $-t=0.65\pm0.05$~GeV$^2$ (red) and $-t=0.95\pm0.05$~GeV$^2$ (blue). Error bars include both statistical and systematic uncertainties as explained in the text.}
\label{fig:final-2}
\end{figure}

\begin{figure}
\vspace{13.cm} 
\includegraphics{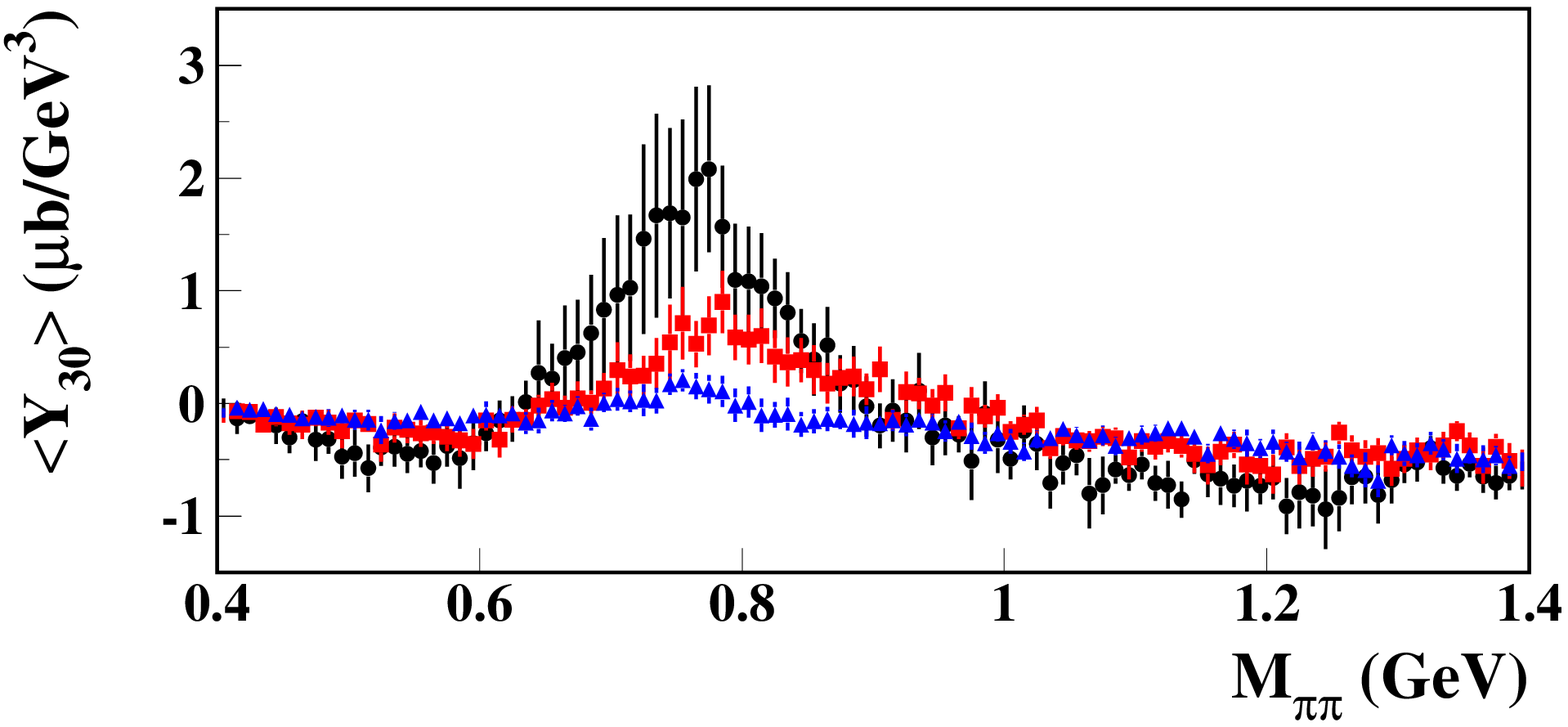}
\includegraphics{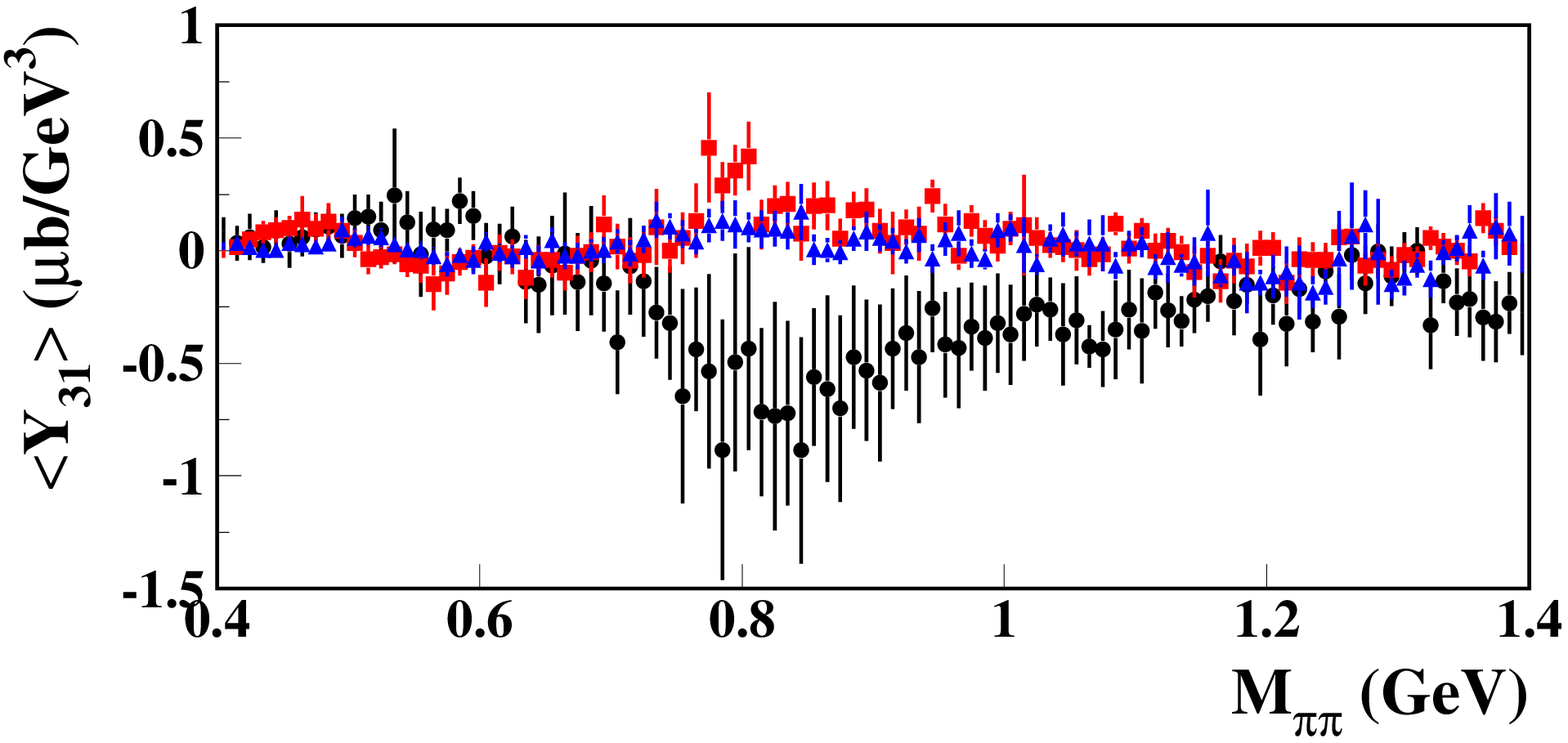}
\includegraphics{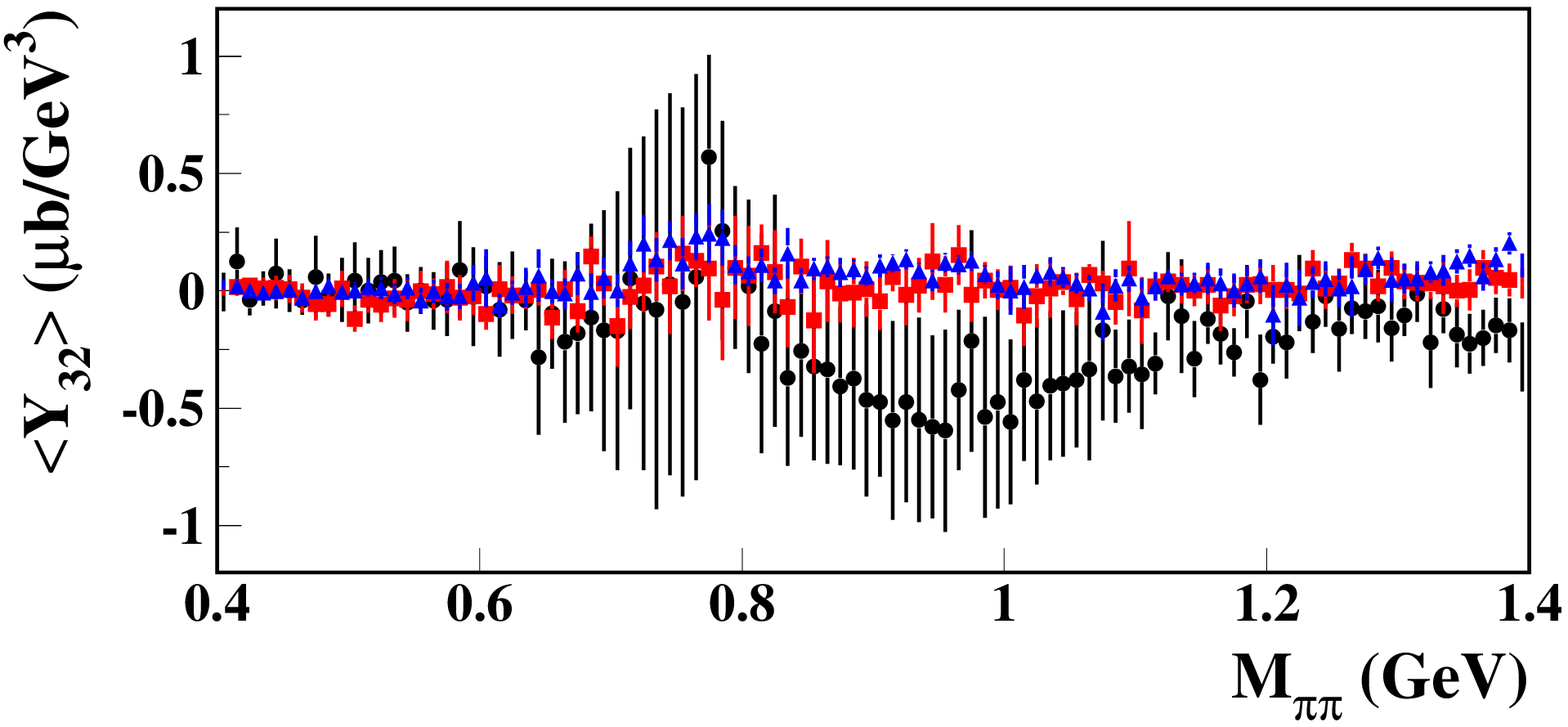}
\caption[]{Moments of the di-pion angular distribution  in $3.2 <E_\gamma< 3.4$~GeV
and $-t=0.45\pm0.05$~GeV$^2$ (black), $-t=0.65\pm0.05$~GeV$^2$ (red) and $-t=0.95\pm0.05$~GeV$^2$ (blue). Error bars include both statistical and systematic uncertainties as explained in the text. }
\label{fig:final-3}
\end{figure}

\begin{figure}
\vspace{13.cm} 
\includegraphics{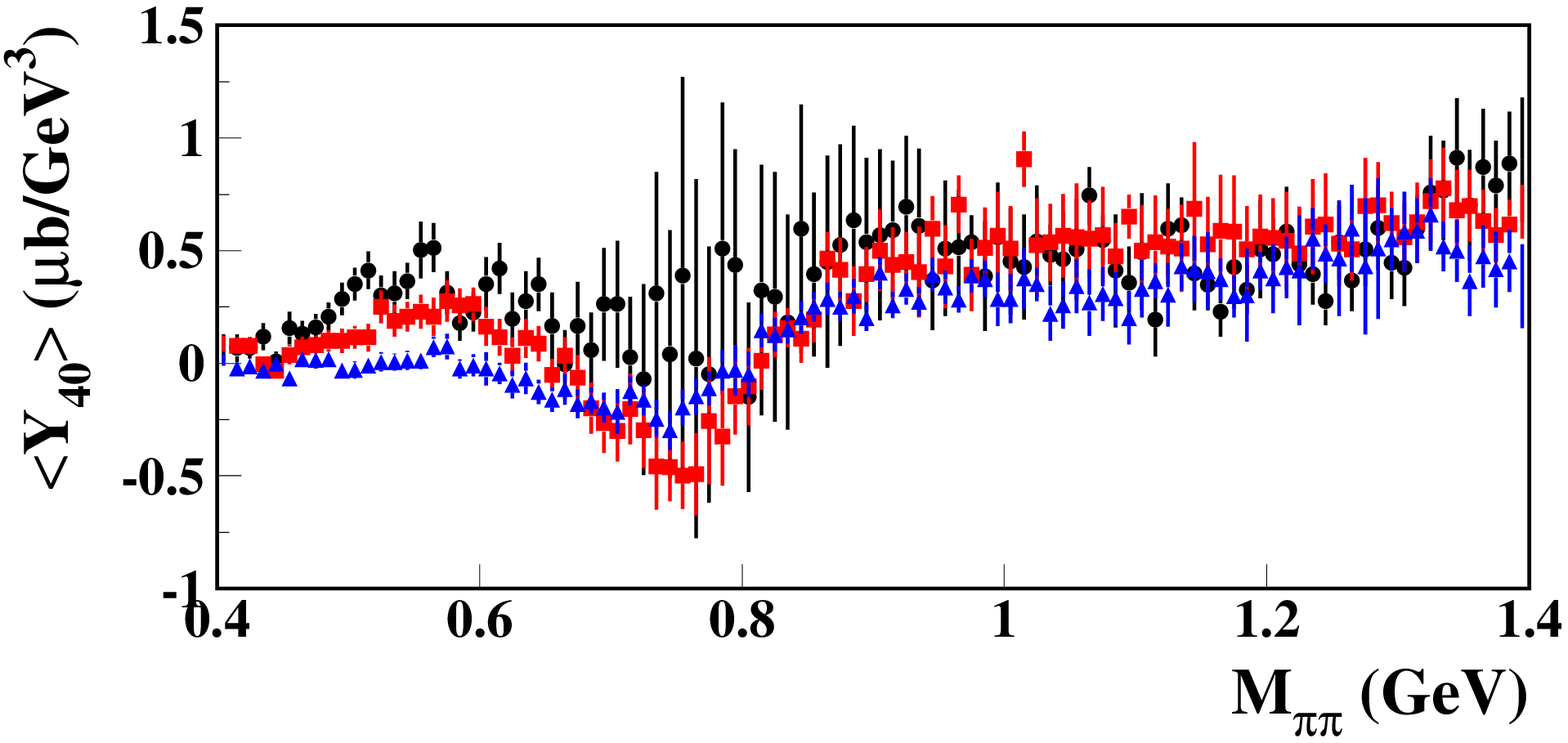}
\includegraphics{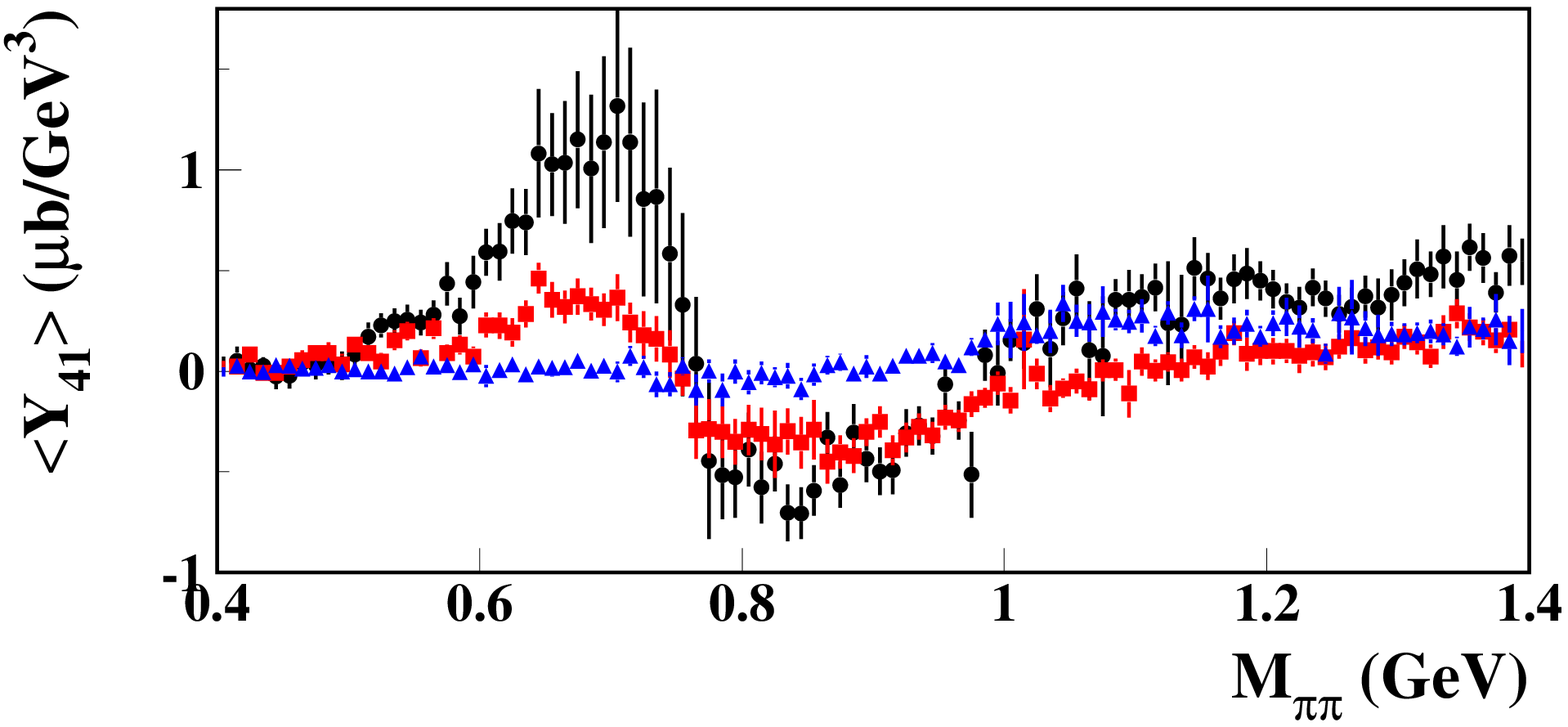}
\includegraphics{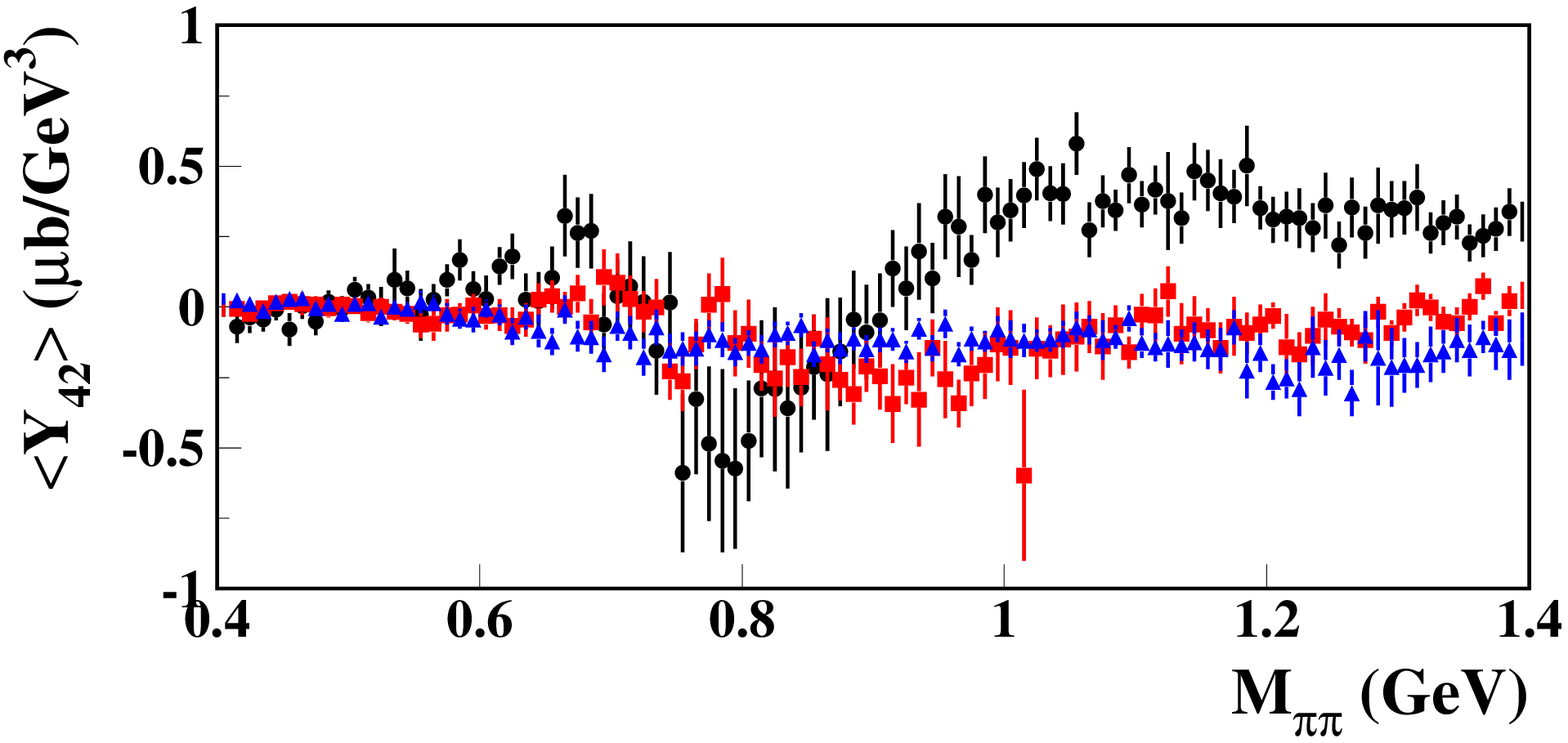}
\caption[]{Moments of the di-pion angular distribution  in $3.2 <E_\gamma< 3.4$~GeV
and $-t=0.45\pm0.05$~GeV$^2$ (black), $-t=0.65\pm0.05$~GeV$^2$ (red) and $-t=0.95\pm0.05$~GeV$^2$ (blue). Error bars include both statistical and systematic uncertainties as explained in the text. }
\label{fig:final-4}
\end{figure}

For most of  the data points, the systematic uncertainties dominate over the statistical uncertainty.
Samples of the  final experimental moments are shown in Figs.~\ref{fig:final-1},~\ref{fig:final-2},~\ref{fig:final-3}, and~\ref{fig:final-4}.
The whole set of moments resulting from this analysis  is  available at the Jefferson Lab~\cite{jlab-db} and the Durham~\cite{dhuram-db} databases. 

As a check of the whole procedure, the differential cross  section $d\sigma/dt$ for the $\gamma p \to p \rho(770)$ meson  has been extracted
by fitting the $\langle Y_{00} \rangle$ moment in each $-t$ bin with a Breit-Wigner plus a first-order polynomial background. 
The agreement within the quoted uncertainties  with a previous CLAS measurement~\cite{Battaglieri},
as well as the world data~\cite{ABBHHM}, gives us confidence in the analysis procedure.

\section{\label{sec:disp} Partial wave analysis}

In the previous section we discussed how moments of the angular distribution of the $\pi^+\pi^-$ system, 
$\langle Y_{LM} \rangle$,  were extracted from the data in each bin in photon  energy, momentum transfer and di-pion mass. 
In this section we describe how partial waves were parametrized and extracted by fitting the experimental moments.

Moments can be   expressed as bi-linear in terms of the amplitudes 
$a_{lm}=a_{lm}(\lambda,\lambda',\lambda_\gamma,E_\gamma,t,M_{\pi\pi})$ with  angular momentum $l$ and $z$-projection $m$
(in the chosen reference system $m$ coincides with the helicity of the di-pion system) as:
\begin{equation}
\langle Y_{LM} \rangle =
\sum_{l'm',lm , \lambda, \lambda'} C(l'm',lm, LM) \times  a_{lm} \;  a^*_{l'm'}\label{Ytheor},
\end{equation}
where $C$  are Wigner's {\it 3jm}  coefficients, $\lambda_\gamma$ is the helicity of the photon,
and $\lambda$ and  $\lambda'$ are  the initial and final nucleon 
helicity, respectively. The explicit forms of the moments with  $L \le 4$ in terms of amplitudes  
with $l=0$ ($S$-wave), $l=1$ ($P$-wave), $l=2$ ($D$-wave), and  $l=3$ ($F$-wave)
are given in Appendix~\ref{app:A}.

\subsection{Helicities, isospin and coupled-channels dependence}
The photon helicity was  restricted to  $\lambda_\gamma  = +1$
since the other amplitudes are related by parity conservation.
In addition, some  approximations in the parametrization  of the partial 
waves were adopted to reduce the number of free parameters in the fit
and are discussed below.

\begin{itemize}
\item The number of waves  was reduced restricting the analysis to  $|m| \le 1$ since 
$m=2$ is only possible for
$l\ge 2$ ($D$ and $F$ waves), which are expected to be small  in the mass range considered~\cite{Ballam_1,Ballam_2}.
In the chosen reference system, $m$ coincides with the helicity of the di-pion system and, since we used as
a reference the wave with  $\lambda_\gamma  = +1$, the three  values of  $m$ have a simple
interpretation in terms of  helicity transfer from the photon to the $\pi\pi$-system:
$m = +1$  corresponds to the non-helicity-flip amplitude ($s$-channel helicity conserving)
that is expected to be  dominant~\cite{Ballam_2}, while $m = 0, -1$  correspond 
to one and two units of helicity flip, respectively. 
In the case of the $S$-wave ($l=m=0$), only one amplitude is considered.

\item
The dependence on the nucleon helicity was simplified as follows.  For a given $l,m,E_\gamma,t$ set,  there are   four independent
 partial wave amplitudes corresponding to the four combinations of initial  and final nucleon helicity, $\lambda$ and $\lambda'$.
It is expected that the dominant amplitudes require no nucleon helicity flip~\cite{Ballam_2}.
Without nucleon polarization information it is not possible to extract all four amplitudes. Thus our strategy  is to consider
in the analysis only the dominant ones or to exploit possible relations among them.
For example, in the Regge  $\rho$ and $\omega$ exchange model, the following
relations are satisfied by the $S$-wave amplitudes: $(\lambda\lambda') = (++) = (--)$ and $(+-)= - (-+)$,
where $\pm$ corresponds to helicity $\pm 1/2$. 
More generally,  by examining the experimental moments,   we observe that
the interference between the dominant $P$-wave, 
seen in the $\langle Y_{21} \rangle$ moment
in the $\rho$ region,
indicates that the $P_{m=+1}$ and the $P_{m=0}$ amplitudes  are out of phase.
For a single nucleon-helicity amplitude,  this would imply a difference between the $\langle Y_{11}\rangle$ and
$\langle Y_{10} \rangle$ moments, arising primarily  from the interference between
the $S$-wave and the $P_{m=+1}$ and $P_{m=0}$
waves, respectively, in the $\rho$ region where the $S$ amplitude  does not vary substantially.
The data suggests, however, that  both $\langle Y_{11} \rangle$ and $\langle Y_{10} \rangle$ peak near the position of the $\rho$.
A possible explanation for the behavior of the  data   is the following:
the dominant $P_{m=+1}$ amplitude may originate from the  helicity-non-flip diffractive process and the $P_{m=0}$ amplitude
from a nucleon-helicity-flip vector exchange, which is also expected to contribute to the  $S$-wave production.
This would also  explain why the  $\langle Y_{11} \rangle $ and $\langle Y_{10} \rangle$ moments have comparable magnitudes.
To  accommodate such behavior, at least two nucleon-helicity amplitudes are required.
In addition, since strong interactions conserve isospin, it is convenient  to write the $\pi\pi$
amplitudes in the isospin basis. Each amplitude  was then expressed as a
linear combination  of $\pi\pi$ amplitudes of fixed isospin $I$ (with $I=0,1,2$).

\item The coupling of the $\pi \pi$ system to other channels was taken into account 
introducing a multi-dimension channel space:
for a given isospin $I$ in the partial wave $l$, the amplitudes depend also on  an index $\alpha$ that  runs over different 
di-meson systems.  For example,  $\alpha=1$ 
corresponds to $\pi\pi$, $\alpha = 2$ to $K{\bar K}$, $\alpha=3 $ to $\eta\eta$, etc. 
In the subsequent analysis we will restrict the channel space to include the $\pi\pi$ and $K{\bar K}$ channels,
which are the only channels relevant in the energy range considered. 

According to these considerations, the moments were fitted to a set of amplitudes  given by:
\begin{equation}
a^{I,\alpha}_{lm,i}(E_\gamma,t,M_{\pi\pi}) 
\end{equation}
for each $l,m$,  $|m|\le 1$, with $i=1,2$ corresponding
to the nucleon helicity non-flip and helicity-flip of one unit, isospin $I=0,1,2$
and channel  $\alpha$.
\end{itemize}

\begin{figure}
\vspace{13.cm}
\includegraphics{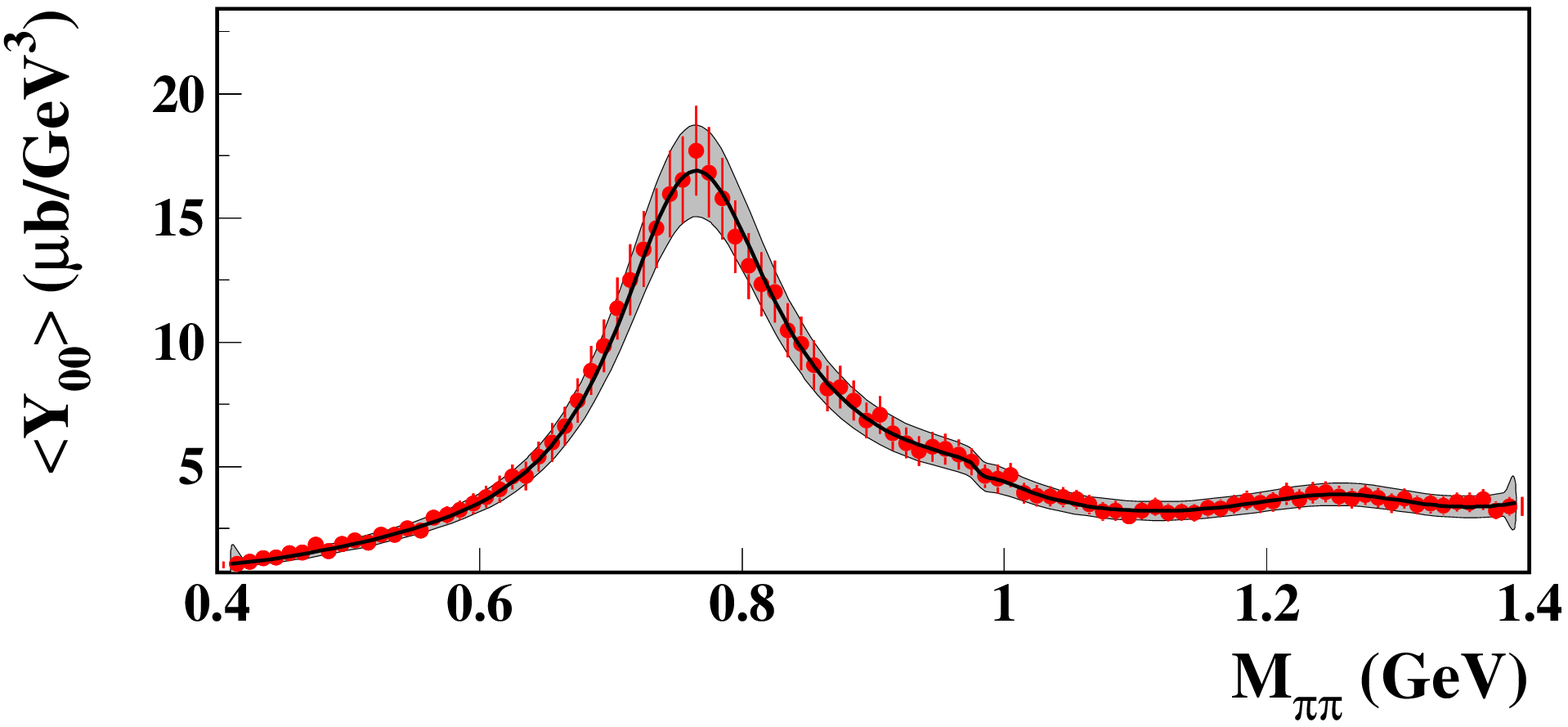}
\includegraphics{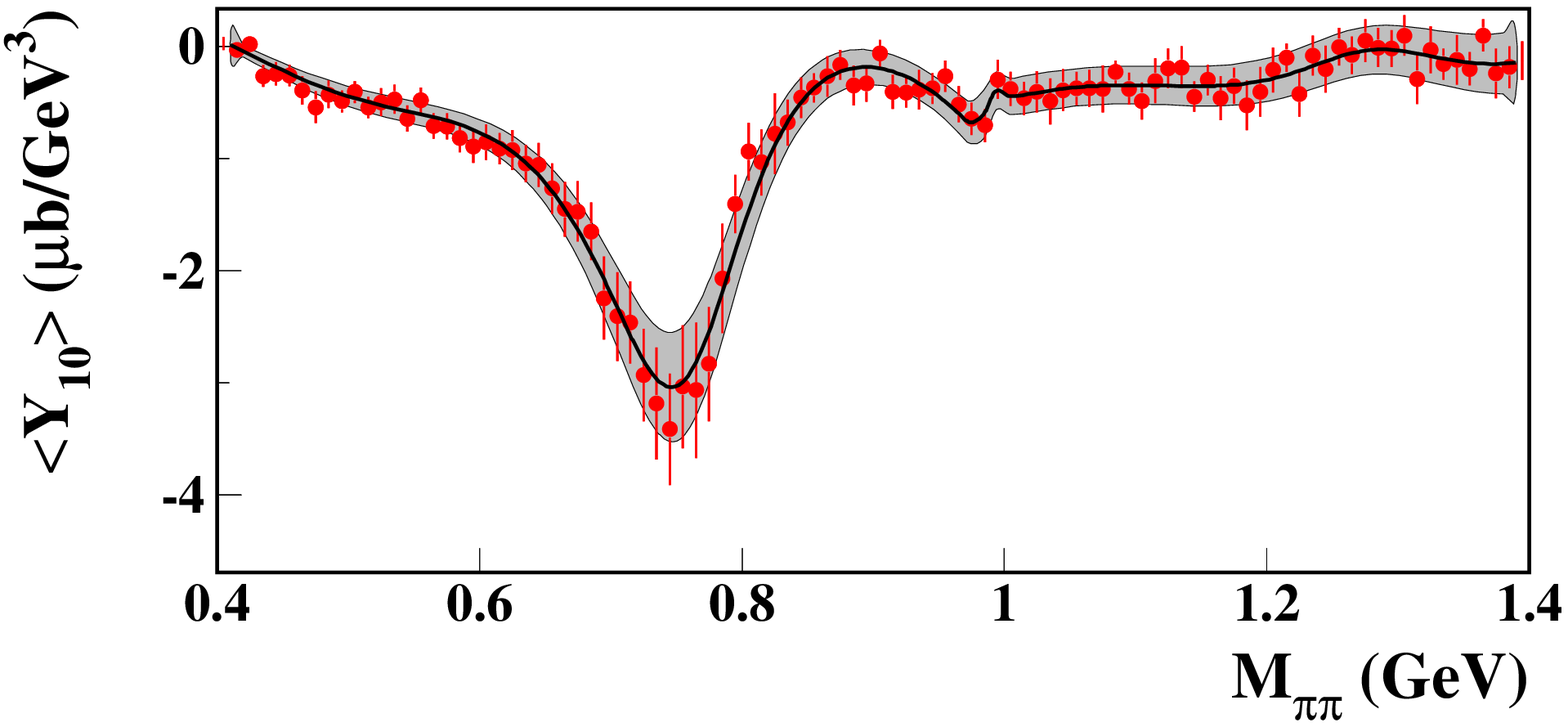}
\includegraphics{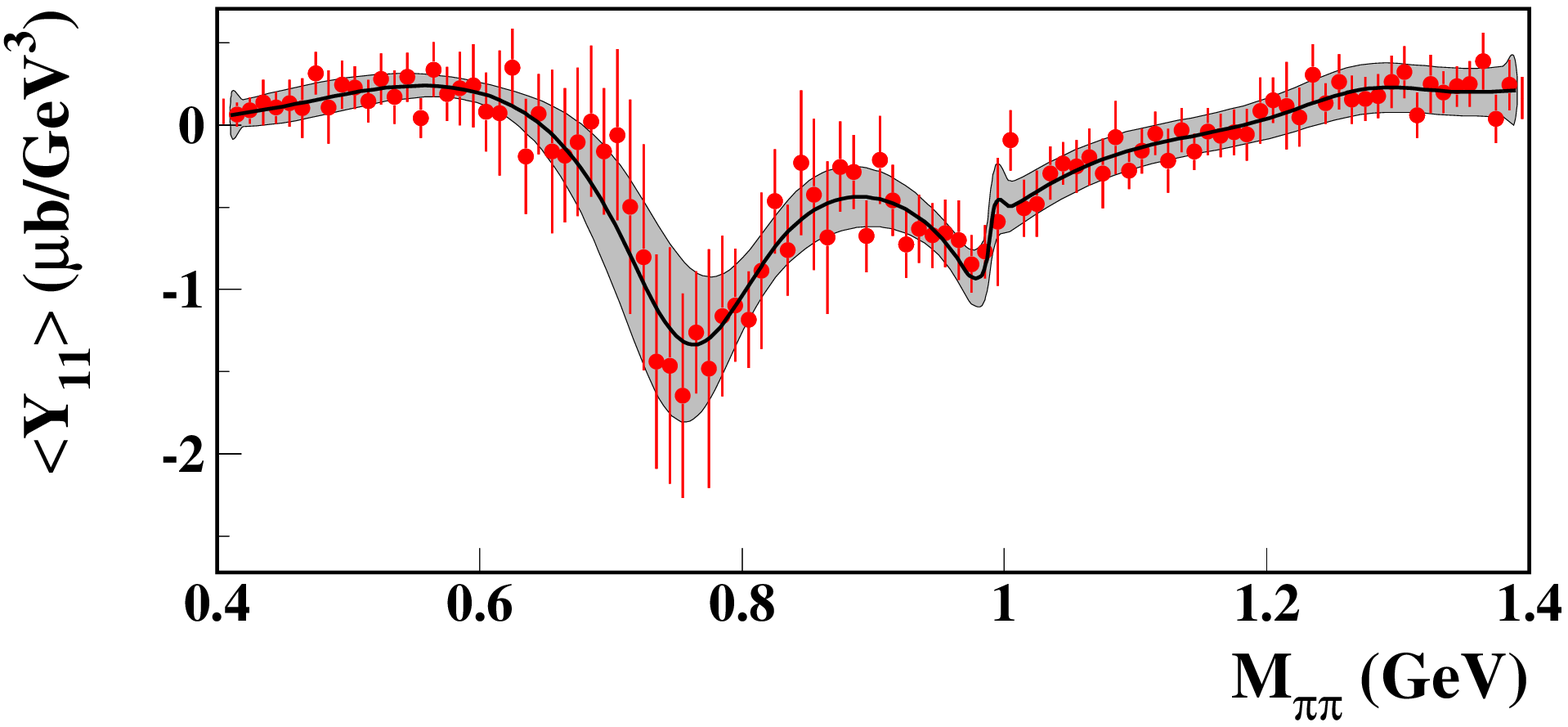}
\caption[]{Fit result (black line) of the final experimental moments (in red) for  $3.2 <E_\gamma< 3.4$~GeV  and $0.5<-t<0.6$~GeV$^2$. 
The systematic  uncertainty and fit  uncertainty are
added in quadrature and are shown by the gray band.}
\label{fig:finalfit-1}
\end{figure}

\subsection{Amplitude parametrization} 
\label{anal}

For each helicity  state  of the target $\lambda$, recoil nucleon $\lambda'$, and  $\pi\pi$ system $m$,  in a given $E_\gamma$ and $t$ bin, 
the corresponding  helicity amplitude  $a_{lm}(s=M^2_{\pi\pi})$, was expressed using a dispersion 
relation~\cite{Aitchison:1976nk,aitchison-1,aitchison-2,ascoli,goradia,Bowler:1975my,Basdevant:1976jg} as follows:
\begin{eqnarray}
\label{disp}
a_{lm,I}(s) & = & {1\over 2} [ I  + S_{lm,I}(s) ] \tilde a_{lm,I}(s)\\ \nonumber
& - &  {  1\over {\pi}} D_{lm,I}^{-1}(s) PV  \int_{s_{th}} ds' {{ N_{lm,I}(s') \rho(s')  \tilde a_{lm,I}(s')} \over {s'-s}}, 
\end{eqnarray} 
where $PV$ represents the principal value of the integral and $\rho$ corresponds to  the phase space term.
In this expression, 
$I$ and $S_{lm,I}$ are matrices in the multi-channel space ($\pi\pi$, $KK$), as mentioned above.
$N_{lm,I}$ and $D_{lm,I}$ can be written in terms of the scattering matrix of $\pi\pi$ scattering, 
chosen to reproduce the known phase  shifts, inelasticities~\cite{Oller:1998hw,Oller:1998zr},
and the  isoscalar ($l=S,D$), isovector ($l=P,F$) 
and isotensor ($l=S,D$) amplitudes in the range $0.4 \mbox{ GeV} < \sqrt{s}  < 1.4 \mbox{ GeV}$.
Finally, the amplitude $\tilde a_{lm,I}$
represents our ignorance about the production process.

As a function of $s=M^2_{\pi\pi}$,  $a_{lm,I}$ have cuts for $s>4m_\pi^2$ (right-hand cut)
and for $s< m_\pi^2$ (left-hand cut).  
The left-hand cut reflects the  nature of particle  exchanges  determining the  $\pi\pi$
 photoproduction amplitude, 
while the right-hand cut accounts for the final-state interactions  of the produced pions. 
In Eq.~\ref{disp}, these discontinuities are taken into account by the functions $N_{lm,I}$ and $D_{lm,I}$, while  $\tilde a_{lm,I}(s)$ does not have singularities
for $s > 4m_\pi^2$ and can be expanded in a Taylor series:
\begin{equation}\label{taylor}
\tilde a_{lm,I} = \left[{\cal A} + {\cal B} s + {\cal C} s^2 + \cdots \right] [k] 
\end{equation} 
with ${\cal A},{\cal B}, \dots$ being matrices of numerical coefficients to be determined by the  
simultaneous fit of the angular moments defined in  Eq.~\ref{Ytheor} and $[k]= k_{\alpha}^l\delta_{\alpha,\beta}$ used to
take into account the threshold behavior in the $l$-{\it th} partial wave.
All amplitudes but the  scalar-isoscalar are saturated by the $\pi\pi$ state. For the scalar-isoscalar amplitude, the $K\bar K$ 
channel was also included. In addition, to reduce sensitivity to the large energy  behavior of the  ($\pi\pi$,$K\bar K$)  amplitudes,
 the real part of the integral was subtracted and replaced by a polynomial in $s$, whose coefficients   were also fitted.

\begin{figure}
\vspace{13.cm}
\includegraphics{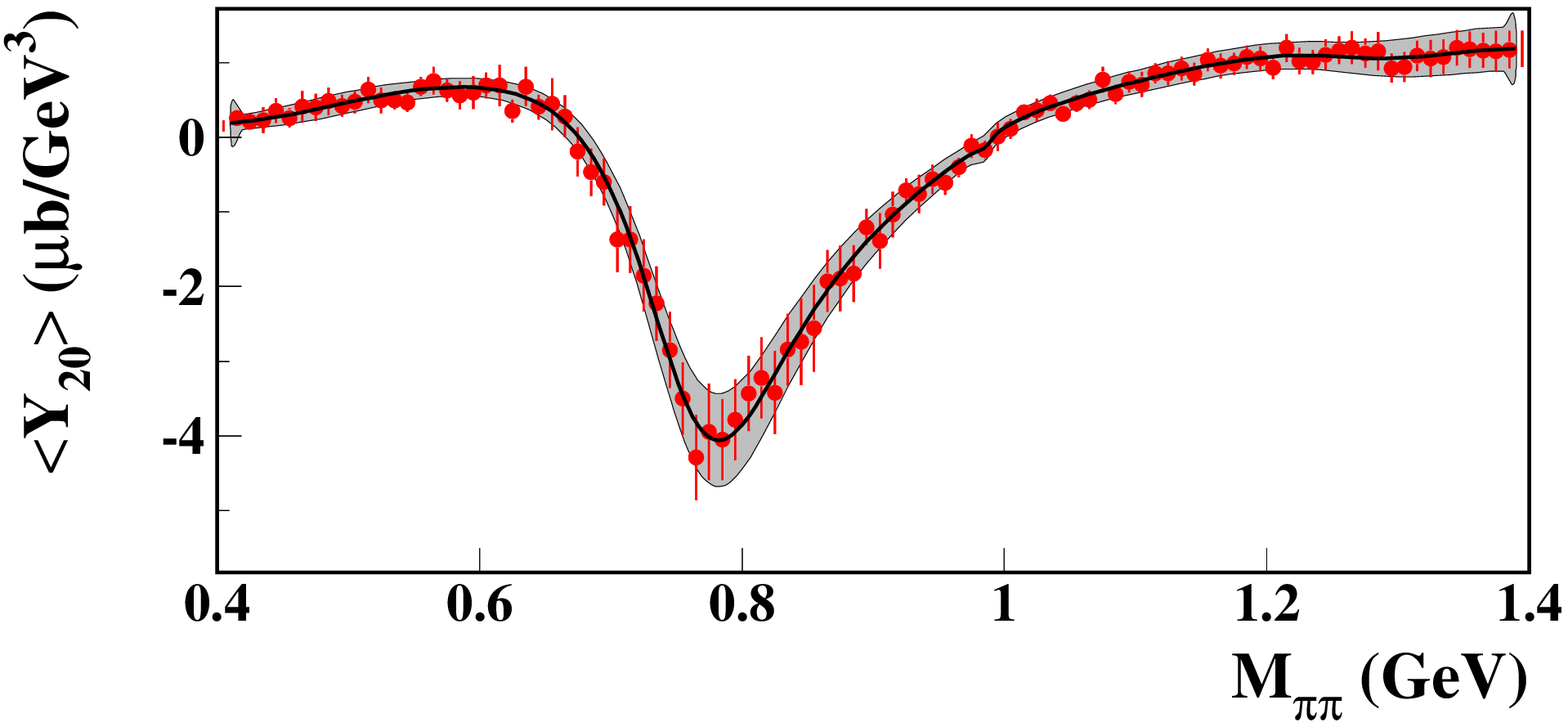}
\includegraphics{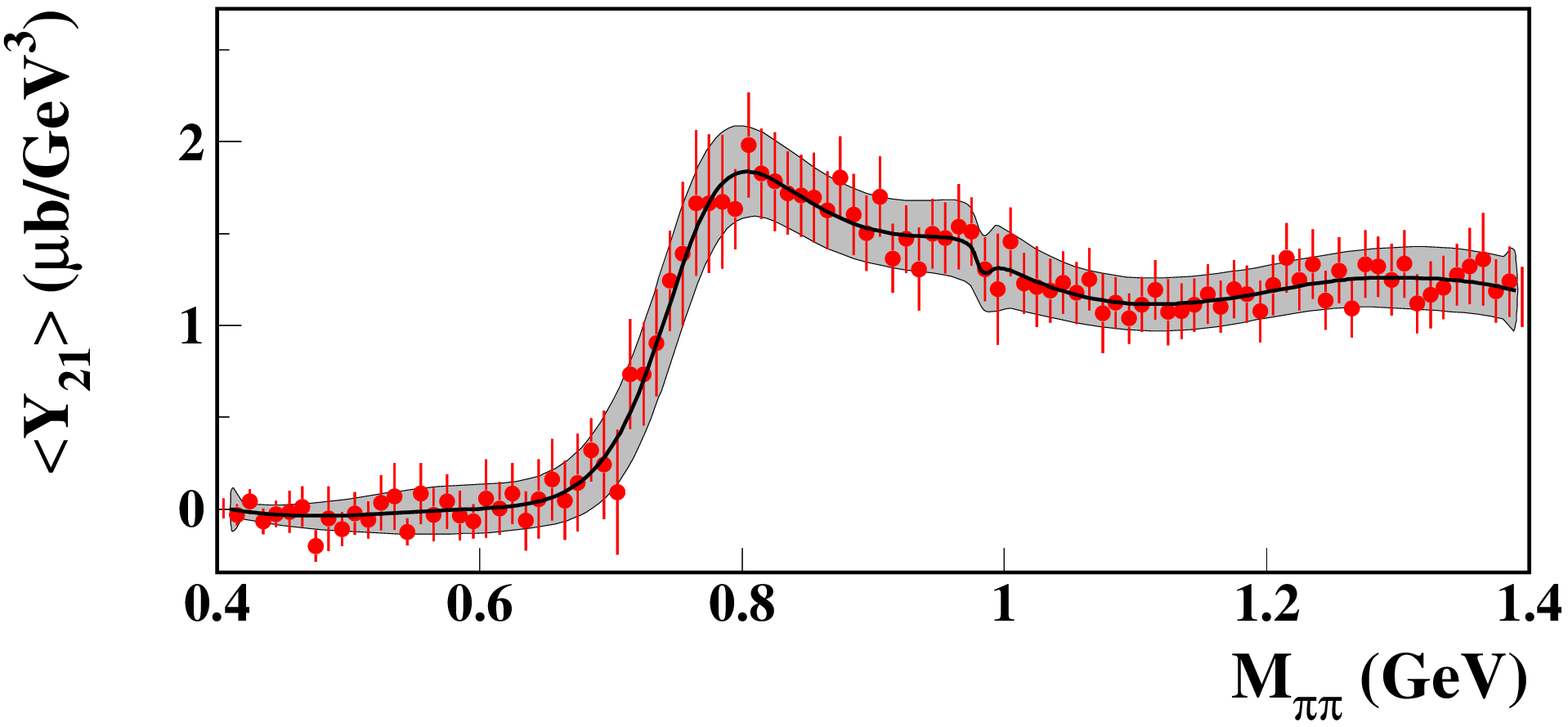}
\includegraphics{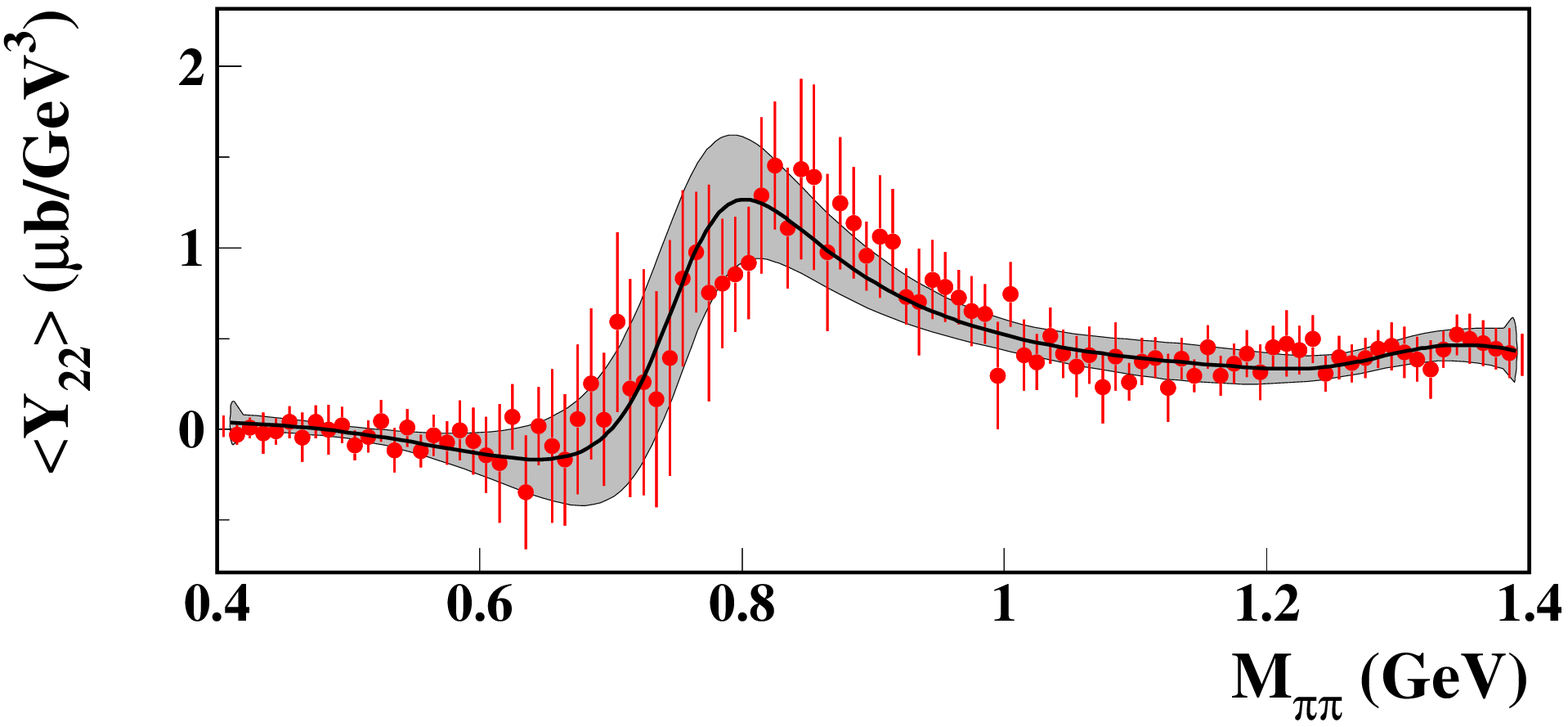}
\caption[]{Fit result (black line) of the final experimental moments (in red) for  $3.2 <E_\gamma< 3.4$~GeV  and $0.5<-t<0.6$~GeV$^2$. 
The systematic  uncertainty and fit  uncertainty are
added in quadrature and are shown by the gray band.}
\label{fig:finalfit-2}
\end{figure}
\begin{figure}
\vspace{13.cm}
\includegraphics{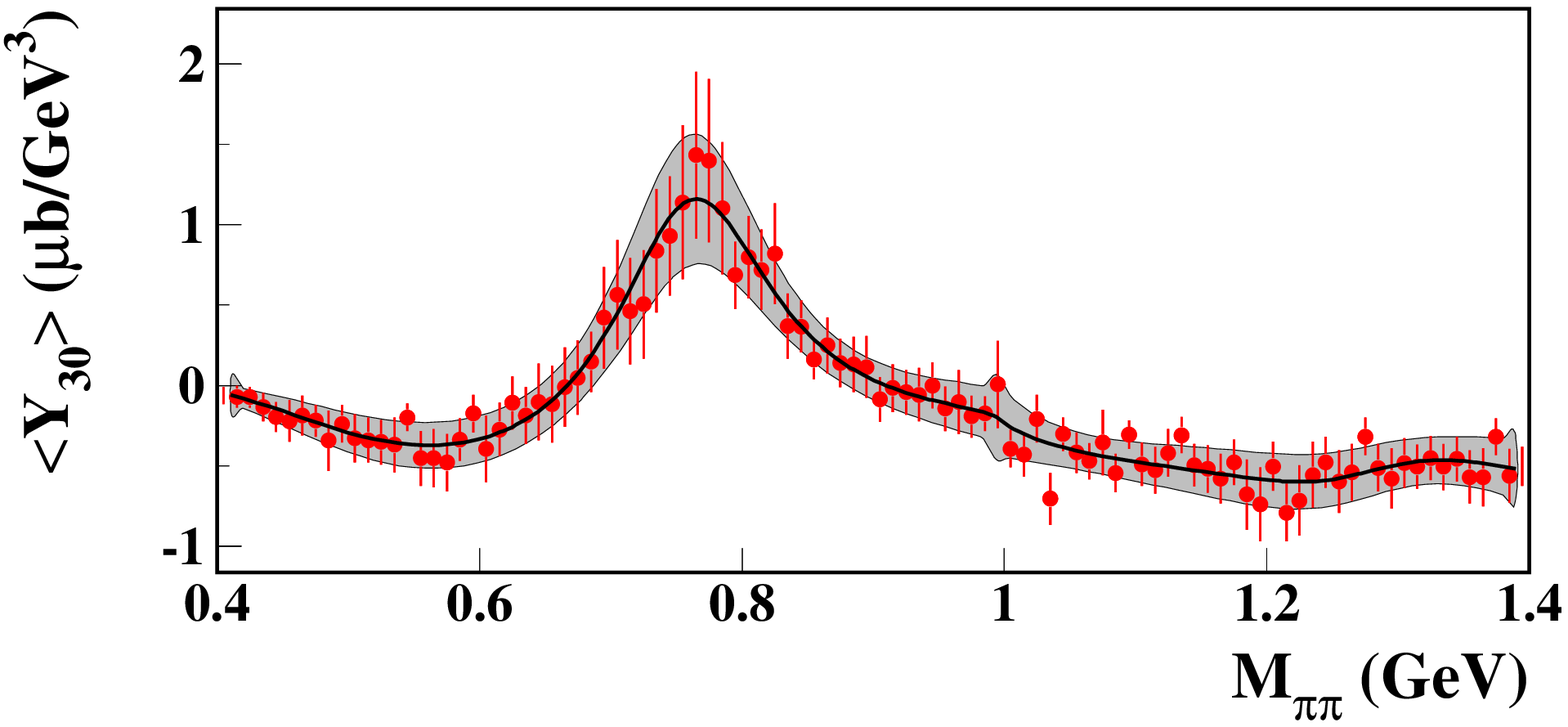}
\includegraphics{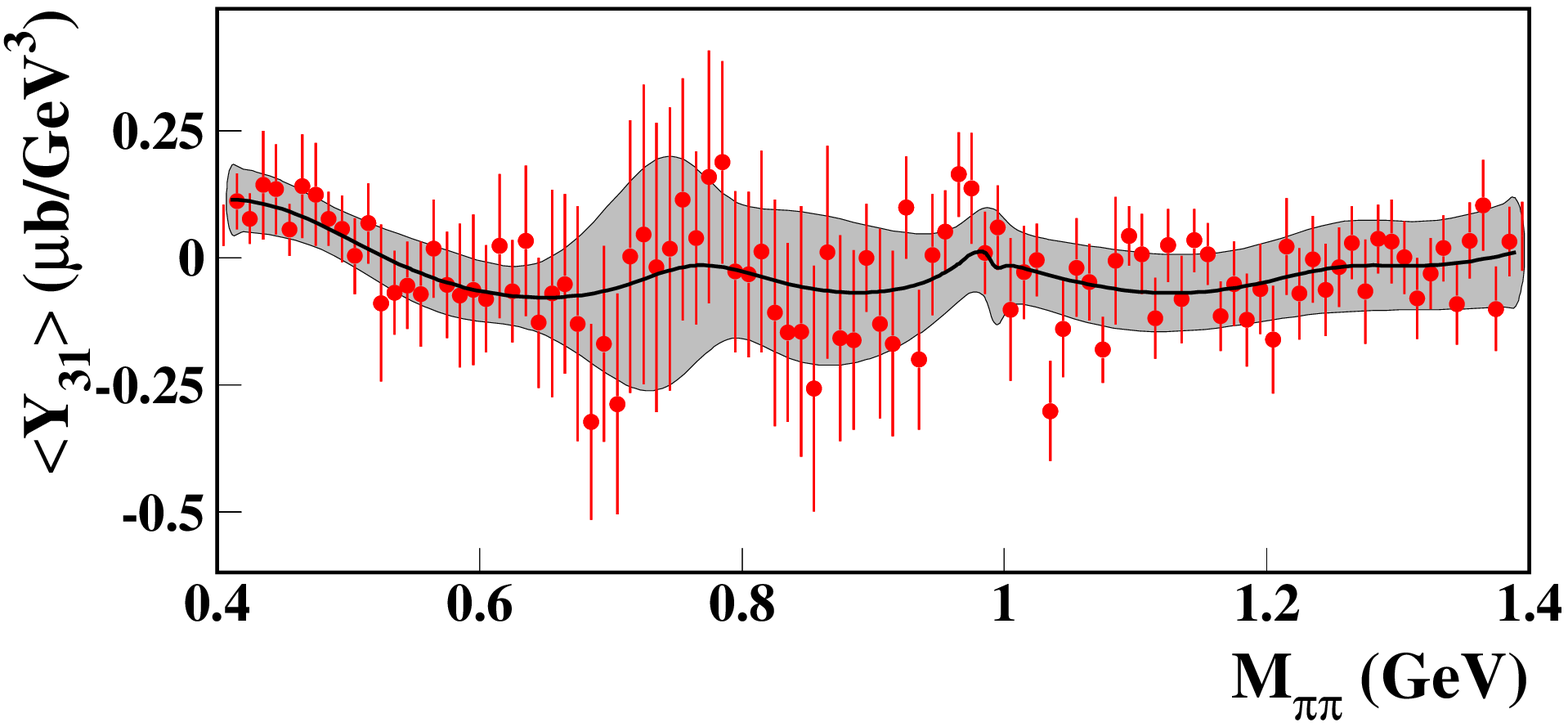}
\includegraphics{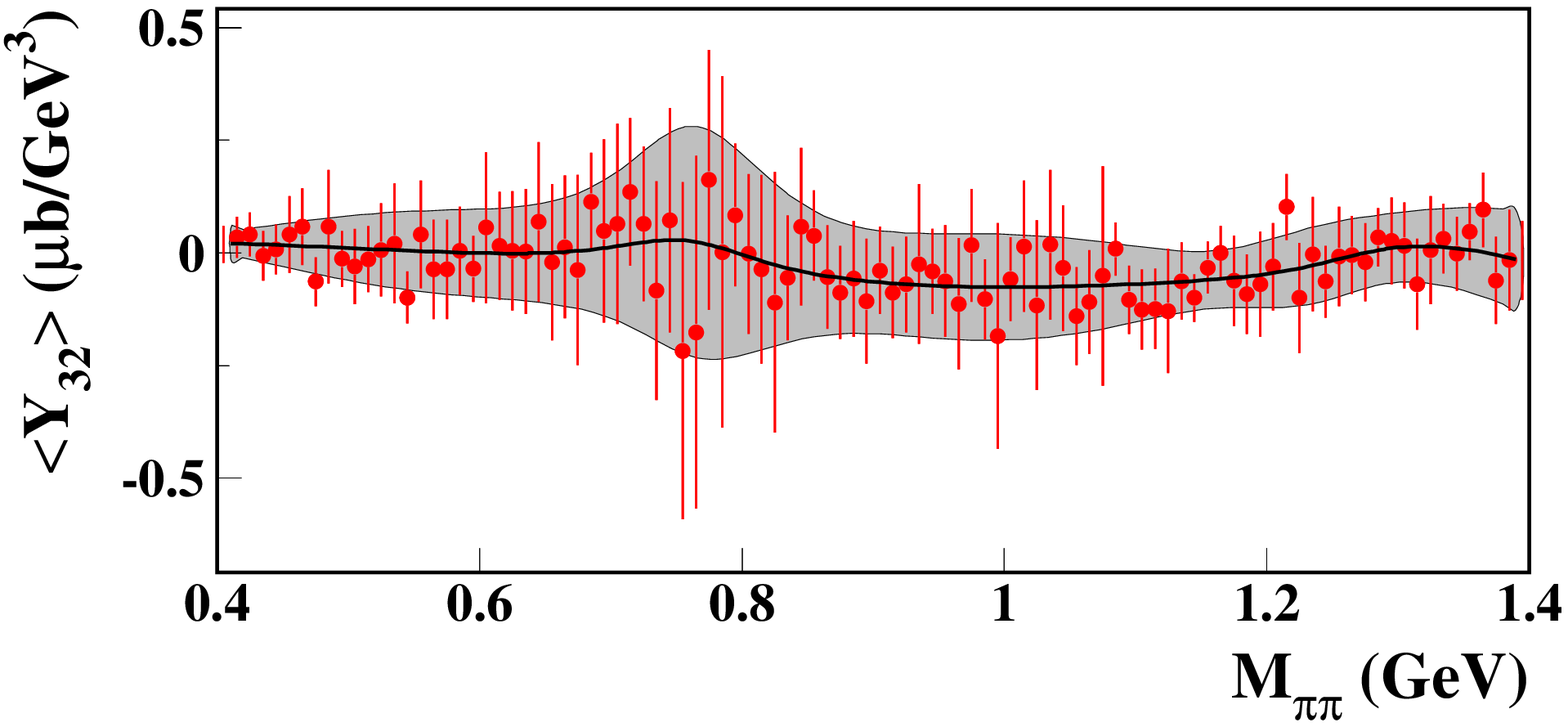}
\caption[]{Fit result (black line) of the final experimental moments (in red) for  $3.2 <E_\gamma< 3.4$~GeV  and $0.5<-t<0.6$~GeV$^2$. 
The systematic  uncertainty and fit  uncertainty are
added in quadrature and are shown by the gray band.}
\label{fig:finalfit-3}
\end{figure}
\begin{figure}
\vspace{13.cm}
\includegraphics{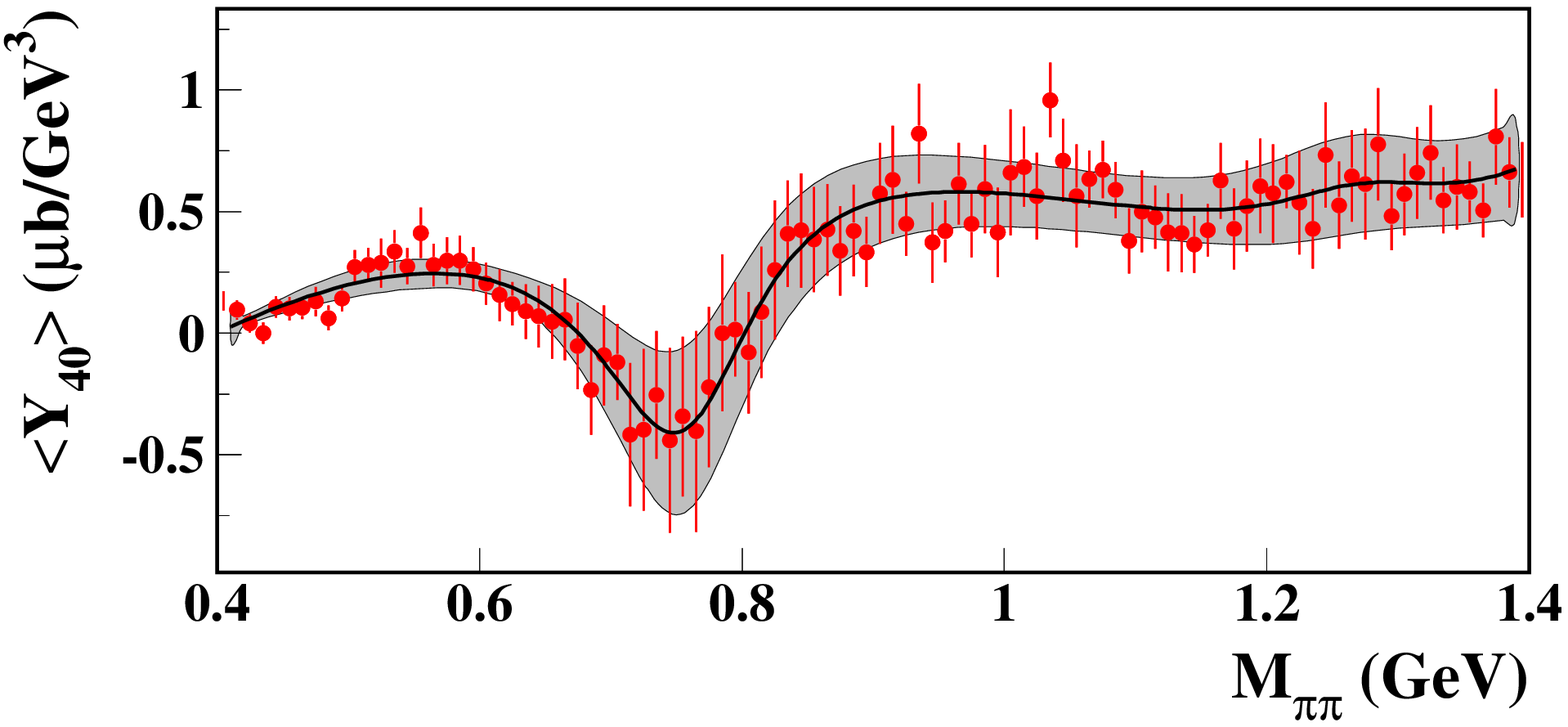}
\includegraphics{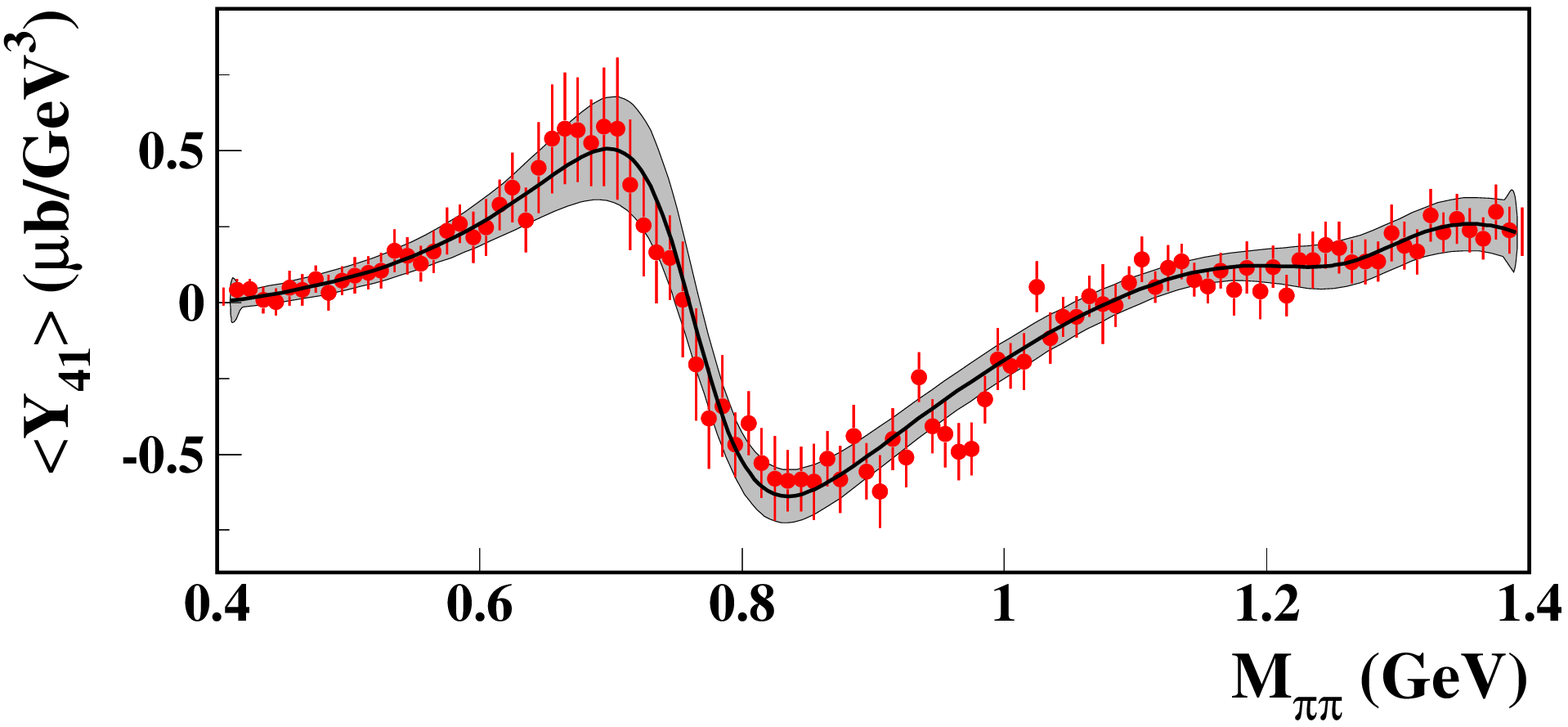}
\includegraphics{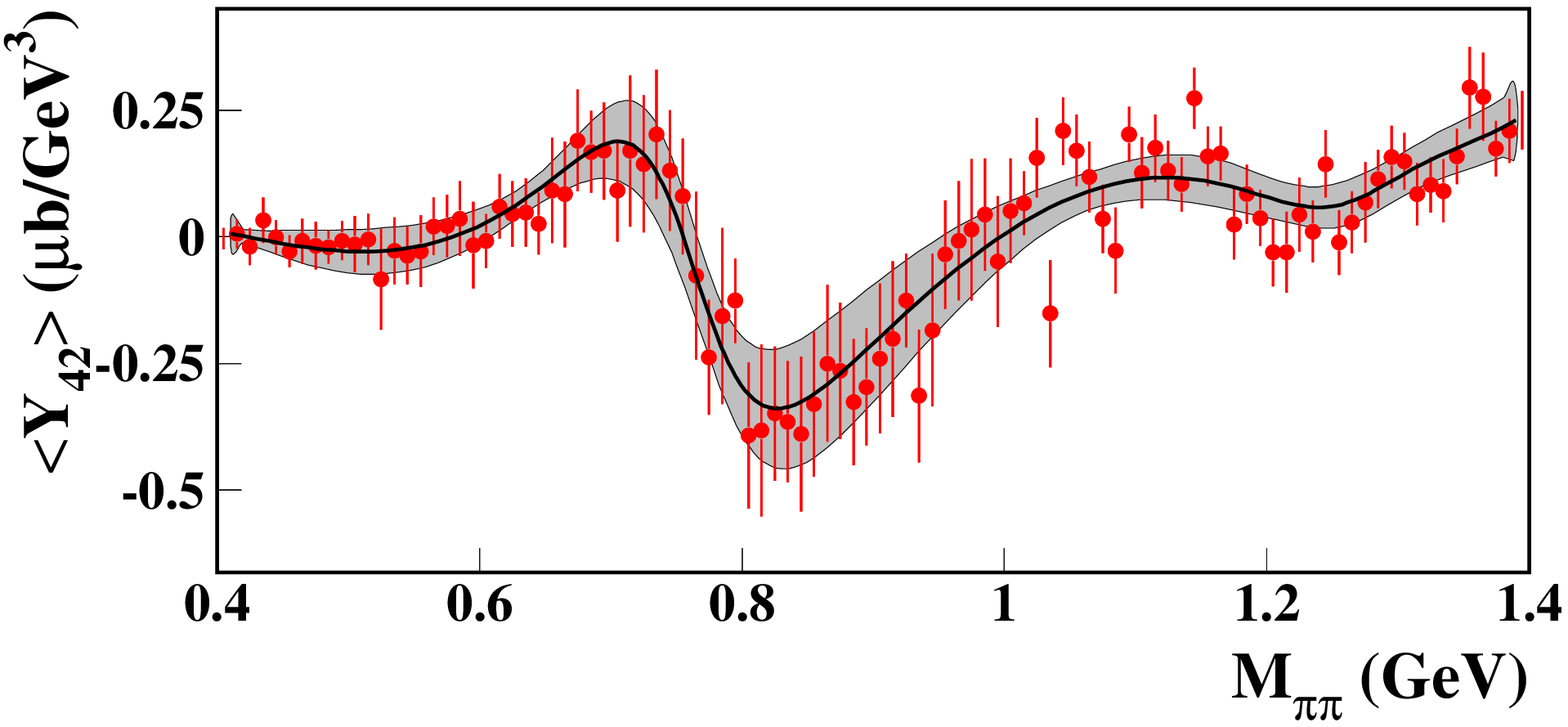}
\caption[]{Fit result (black line) of the final experimental moments (in red) for  $3.2 <E_\gamma< 3.4$~GeV  and $0.5<-t<0.6$~GeV$^2$. 
The systematic  uncertainty and fit  uncertainty are
added in quadrature and are shown by the gray band.}
\label{fig:finalfit-4}
\end{figure}

\section{Results}\label{sec:res}
\subsection{Fit of the moments}
Using the parametrization of the partial waves described in the previous section,
we fitted all moments $\langle Y_{LM}\rangle$ with $L \le 4$ and $M \le 2$ using amplitudes with $l \le 3$ (up to $F$-waves).
In Figs.~\ref{fig:finalfit-1},~\ref{fig:finalfit-2},~\ref{fig:finalfit-3}, and~\ref{fig:finalfit-4}  we present a sample of the fit  results for 
$E_\gamma  = 3.3 \pm 0.1$~GeV and  $0.5<|t|<0.6$~GeV$^2$. 

To properly take into account the statistical and systematic uncertainty contributions  to the experimental moments described in 
Sec.~\ref{par:fin_results}, the four sets of moments resulting from the different fit procedures
were individually fitted and the results were averaged, obtaining the
central value shown by the black line in the figures. 
The error band, shown as a gray area, was calculated following the same procedure adopted for the experimental moments.
The final uncertainty was computed as the sum
in quadrature of the statistical uncertainty  of the fit and the  two systematic uncertainty contributions.
The first is  related to the moment extraction procedure and is evaluated as the variance of the four fit results.
The latter is  
associated with the photon flux normalization and is estimated to be 10\%.
The central values and uncertainties for  all the observables of interest discussed in the following sections were derived 
from the fit results with the same procedure.

The moment $\langle Y_{00} \rangle$, corresponding to the differential production cross section $d\sigma/dtdM$, shows 
the dominant $\rho(770)$ meson peak.
In the $\langle Y_{10} \rangle$ and $\langle Y_{11} \rangle$  moments,  the contribution of the $S$-wave is 
maximum and enters via interference with the $P$-wave. 
In particular the  structure  at $M_{\pi\pi} \sim 0.77\mbox{ GeV}$ in  $\langle Y_{11} \rangle$ 
is due to the interference of the $S$-wave with the dominant, helicity-non-flip wave, $P_{m=+1}$. 
In the  $\langle Y_{10} \rangle$ moment the same structure is due to the 
interference with the $P_{m=0}$ wave, which corresponds to  one unit of helicity flip.
A second dip near $M_{\pi\pi} = 1\mbox{ GeV}$ is clearly visible and 
corresponds to the  production of a resonance that we interpret as the $f_0(980)$. 


\begin{figure}
\vspace{5.4cm} 
\includegraphics{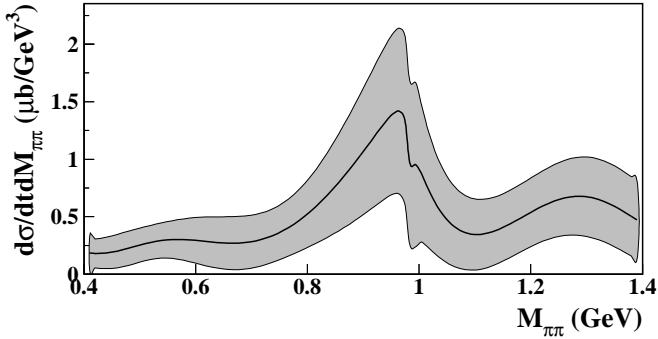}
\caption[]{$S$-wave cross section derived by the fit in the  $3.2 <E_\gamma< 3.4$~GeV  and $0.5<-t<0.6$~GeV$^2$ bin.
The systematic and the fit uncertainties  are  added in quadrature and are shown by the gray band.} 
\label{fig:s-wave}
\end{figure}

\begin{figure}[h]
\vspace{9cm} 
\includegraphics{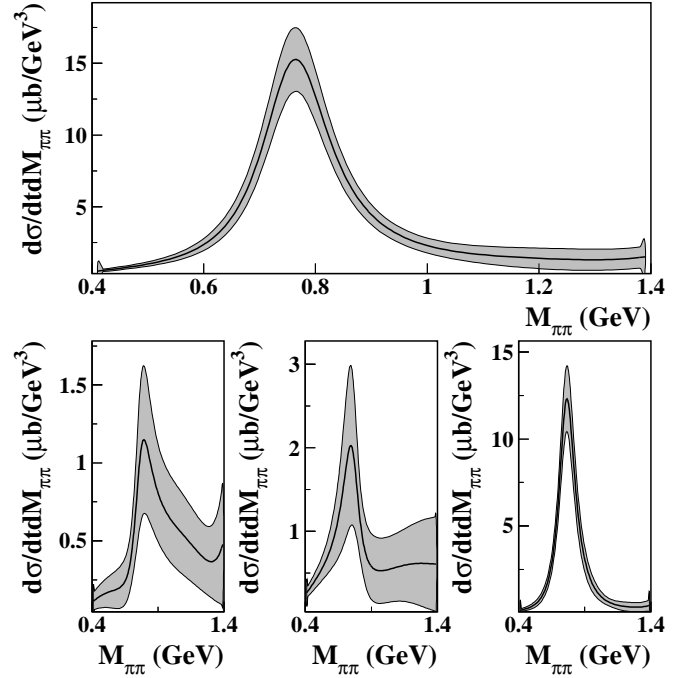}
\caption[]{$P$-wave cross section derived by the fit in the  $3.2 <E_\gamma< 3.4$~GeV  and $0.5<-t<0.6$~GeV$^2$ bin.
Bottom plots:  the same amplitudes for the three possible values of $\lambda_{\pi\pi}$ (from left to right -1, 0 and +1).
The   systematic and fit uncertainties  are added in quadrature and are shown by the gray band.} 
\label{fig:p-wave}
\end{figure}

\subsection{Partial wave amplitudes} 
The square of the magnitude of  the $S$-, $P$-, $D$- and $F$-waves resulting from the fit, summed  over the nucleon spin projections,
is given by:
\begin{eqnarray}
I_{lm} = \sum_{i=1,2} |a_{lm,i}(E_\gamma,t,M_{\pi\pi})|^2. \nonumber \\ 
\end{eqnarray}
When summed over the di-pion helicity, this can be written as:
\begin{eqnarray}
I_{l} =\sum_{m} \sum_{i=1,2}|a_{lm,i}(E_\gamma,t,M_{\pi\pi})|^2, \nonumber \\ 
\end{eqnarray}
where the sum is limited to $m=-1, 0 , 1$ for $l>0$ and to $m=0$ for $l=0$.

\begin{figure}[h]
\vspace{9cm} 
\includegraphics{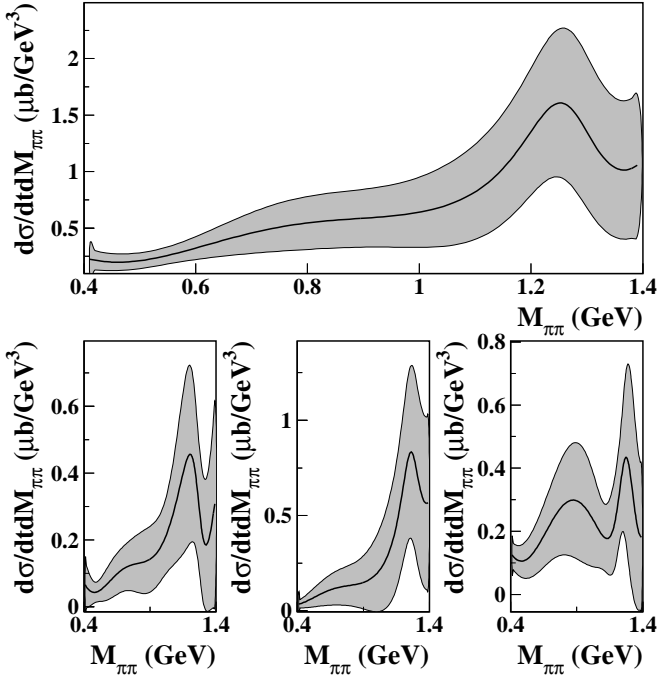}
\caption[]{As Fig.~\ref{fig:p-wave} for $D$-wave.}
\label{fig:d-wave}
\end{figure}

\begin{figure}
\vspace{9cm} 
\includegraphics{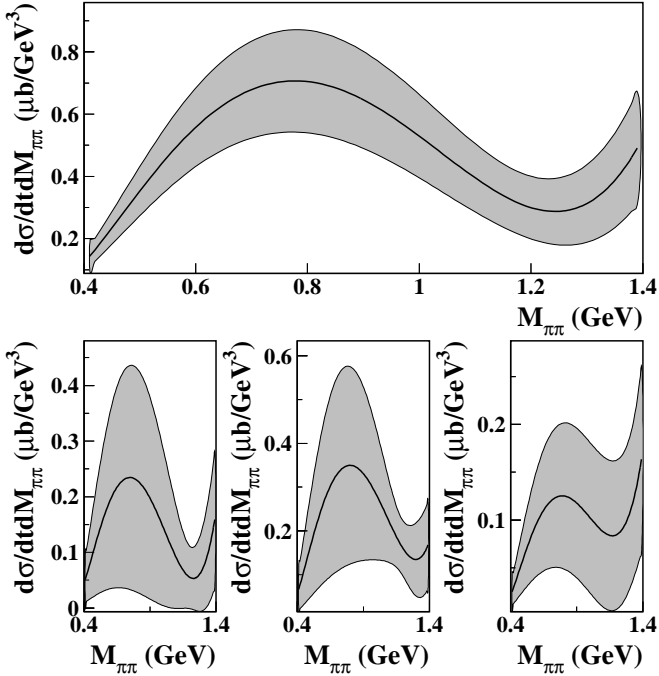}
\caption[]{As Fig.~\ref{fig:p-wave} for $F$-wave.}
\label{fig:f-wave}
\end{figure}

The resulting partial wave cross sections are shown in 
Figs.~\ref{fig:s-wave},~\ref{fig:p-wave},~\ref{fig:d-wave}, and~\ref{fig:f-wave}, for a selected photon 
energy and $-t$ bin. The whole set of partial wave amplitudes resulting from this analysis  is  
available at the Jefferson Lab~\cite{jlab-db} and the Durham~\cite{dhuram-db} databases.

As expected, the dominant contribution from the $\rho$ meson is clearly visible in the  $P$-wave, whose contribution is about 
one order of magnitude larger than the other waves. In particular the main contribution comes from $I_{lm=1,+1}$,
corresponding to a non-helicity flip ($s$-channel helicity conserving) transition.
In the  $S$-wave,  a strong interference pattern shows up around $M_{\pi\pi}=980$~MeV,  which reveals contributions
from the  $f_0(980)$  production.
The contribution from the $f_2(1270)$ tensor meson is apparent in the $D$-wave, while no clear structures are seen in the $F$-wave.

\subsection{Systematic studies}\label{sec:sys}
The error bands plotted in  Figs.~\ref{fig:s-wave},~\ref{fig:p-wave},~\ref{fig:d-wave}, and~\ref{fig:f-wave} include
the systematic uncertainties related to the moment extraction  and the photon flux normalization
as discussed in  Sec.~\ref{par:syserrmom}. In addition, for the $S$-wave, where the $f_0(980)$ contribution 
is strongly affected by interference, detailed systematic studies using both Monte Carlo and data were performed.

\begin{figure}
\vspace{9cm} 
\includegraphics{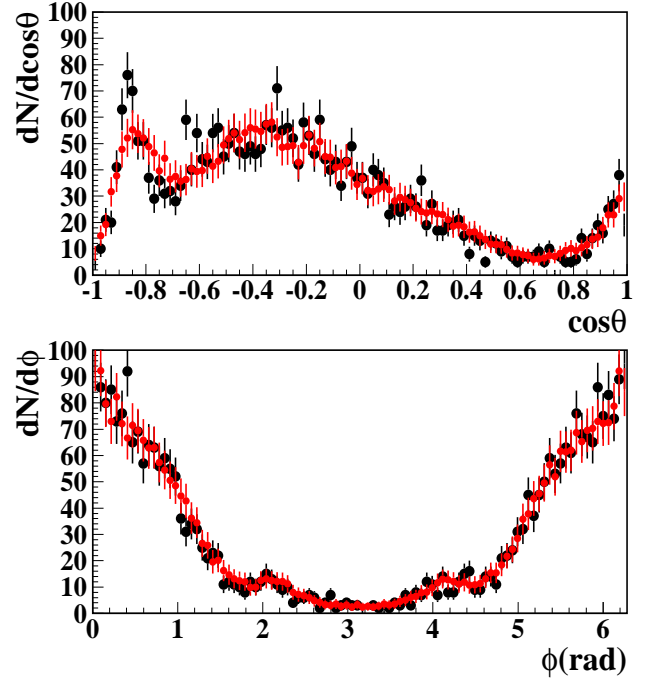}
\caption[]{Pion angles in the $\pi^+ \pi^-$  helicity rest frame 
for $M_{\pi \pi}$ in  the $f_0(980)$ mass region ($M_{\pi \pi}= 0.985\pm 0.01$~GeV).
Experimental data are plotted  in black and fit results in red.}
\label{fig:sys-hel}
\end{figure}

\begin{figure}
\vspace{12cm}  
\includegraphics{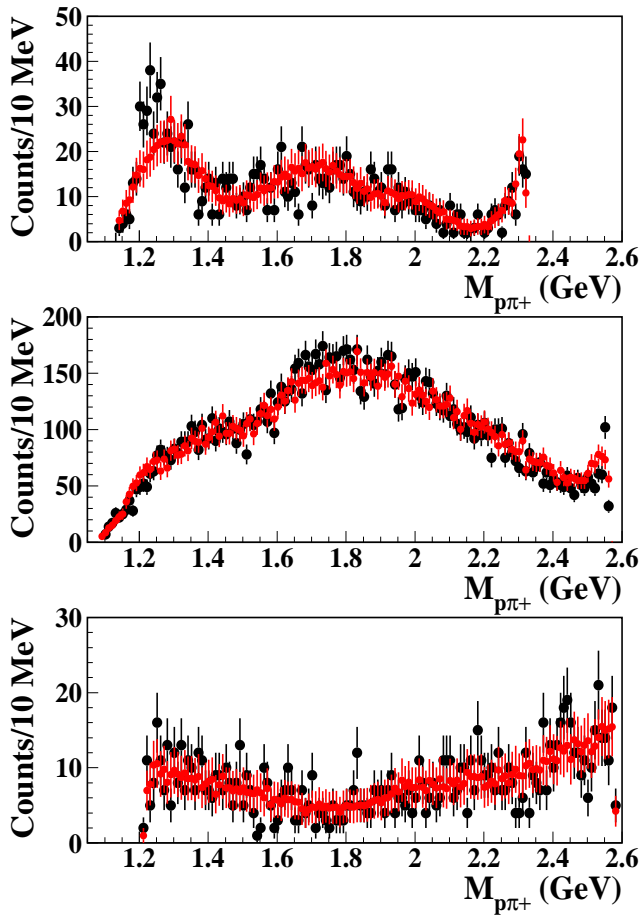}
\caption[]{$M_{p\pi^+}$ distribution in three different  $M_{\pi \pi}$ mass regions 
(bottom: $M_{\pi \pi}=0.475 \pm 0.01$~GeV, middle: $M_{\pi \pi}=0.775 \pm 0.01$~GeV, top: $M_{\pi \pi}=1.295 \pm 0.01$~GeV).
Experimental data are plotted  in black and fit results in red.}
\label{fig:sys-ppi+}
\end{figure}

In order to test the approximation introduced by the  truncation to $\l_{max}$=4 in the moment extraction, we first
verified the fit was able to reproduce the  experimental distributions in the kinematic range of interest.
Figure~\ref{fig:sys-hel} shows the comparison between data and fit results for the  decay angles in the helicity system 
with $M_{\pi \pi}$ in  the $f_0(980)$ mass region 
($M_{\pi \pi}= 0.985\pm 0.01$~GeV). Figure~\ref{fig:sys-ppi+} shows the same comparison for the invariant mass $M_{p\pi^+}$ when three different  
regions of $M_{\pi \pi}$ ($M_{\pi \pi}=0.475  \pm 0.01$~GeV, $M_{\pi \pi}=0.775 \pm 0.01$~GeV, $M_{\pi \pi}=1.295 \pm 0.01$~GeV) were selected.
The good agreement proves  the accuracy of the approximation.

As a second check, we applied the fit 
to pseudo-data  obtained with a realistic
event generator,  processed with the CLAS GEANT-based simulation package and analyzed 
with the same procedure used for the data. Since the event generator was tuned to previous two-pion photoproduction
measurements, it does not include any explicit limitation on the number of waves.
The reconstructed  moments showed that, with the chosen $\l_{max}$, all fits were capable 
of reproducing the generated moments up to $M_{\pi\pi}\sim 1.1$~GeV. 
Finally, we derived a quantitative estimate of the truncation effect on the $S$-wave squared amplitude
as follows. The results of  a $\l_{max}=8$ fit of the moments was used as input for a new Monte Carlo  event generator.
After being processed in the same way as discussed above, pseudo-data were fitted with $\l_{max}=4$ and the  $S$-wave
amplitude was extracted. The difference between the generated and the reconstructed partial wave cross section was found to be of the order of 
25\% that, added in quadrature to the other systematic uncertainties, was  included in the gray band of Fig.~\ref{fig:s-wave}.

We also demonstrated that no structures similar to the 
narrow interference pattern we are interpreting as the evidence of the  $f_0(980)$ were
created by distortions induced by the CLAS acceptance. This check was performed generating events after removing the 
$f_0(980)$ contribution, and verifying that no spurious structures appeared in the spectra after the 
full GEANT simulation and reconstruction. 

\begin{figure}
\vspace{10cm}  
\includegraphics{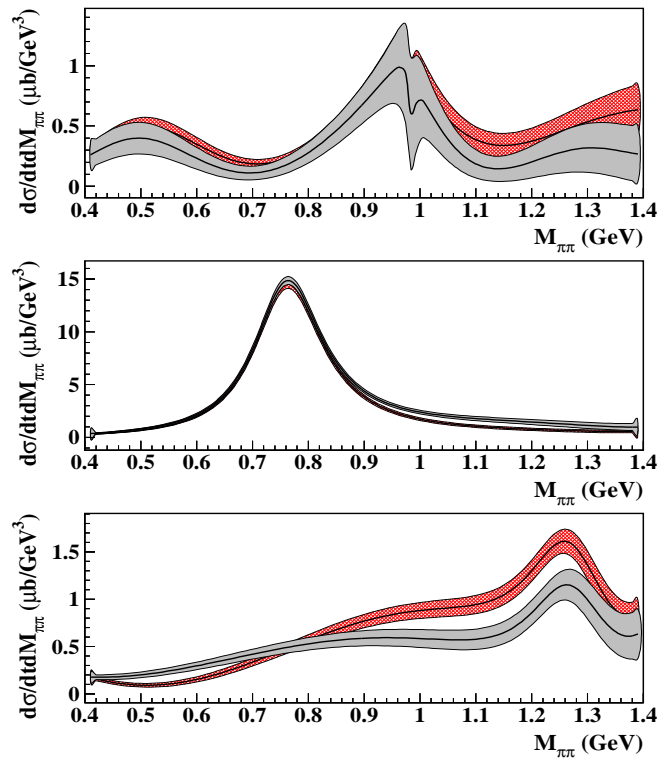}
\caption[]{$S$-, $P$- and $D$-wave cross sections in the  $3.4 <E_\gamma< 3.6$~GeV  and $0.5<-t<0.6$~GeV$^2$ bin. 
The gray and red bands show the results of the standard fit and of the fit performed adding a second-order 
polynomial to the partial wave expansion of the moments to account for the baryon resonance contributions.
The width of the bands represents the fit uncertainties only.
Fit results are shown for a specific parametrization of the moments (second method, see Sec.~\ref{par:syserrmom}).
}
\label{fig:background}
\end{figure}

In addition, the effects of baryon resonance contributions to the di-pion mass spectrum were studied performing the fit of the 
moments with the inclusion of an incoherent background. In fact, the background in the di-pion mass spectrum introduced by the reflection
of the baryon resonances is expected to be smooth and structureless, contributing to all waves. Therefore this was parametrized as 
 a second-order polynomial in $M_{\pi\pi}$ that was summed 
to the parametrization of the moments in terms of 
partial waves used in the standard fits. 
From this study we concluded that the background contribution is small, smooth and does not affect the quality of the fit. 
The comparison of the fit results with and without the inclusion of this additional  background  indicates that
the $P$-wave and the $S$-wave in the $f_0(980)$ region are only slightly  affected, as shown in Fig.~\ref{fig:background}.
On the contrary, the low mass $S$-wave, corresponding
to the $\sigma(600)$ region,  and the $D$-wave,  corresponding to the $f_2(1270)$ region, show a
significant variation and, therefore, a more complete analysis should be performed to
extract reliable information in these mass ranges.
A similar conclusion was drawn by comparing the analysis results excluding the $\Delta(1232)$, 
the dominant baryon resonance contribution for this final state,  
with the cut $M(p\pi^+)>1.4$ GeV. A negligible effect was found on the rapid motion around the narrow 
$f_0(980)$ meson, while a larger variation was observed at higher  values of the  $M(\pi\pi)$ mass. 

To verify the  stability of the fit of moments  in the region of the  $f_0(980)$, the whole analysis was repeated reducing the
$M_{\pi \pi}$ bin size from 10 to 5 MeV. The results obtained in the two cases were 
found to be consistent.

As a final check, the sensitivity to the specific choice of the number of terms used  in the Taylor expansion of the
amplitudes $\tilde a^L$ (see
Eq.~\ref{taylor}) was tested performing the partial wave analysis fits both with a  second- and fourth-order polynomial.  
The effect was found to be negligible compared to the other systematic uncertainties.






\begin{figure} 
\vspace{10.cm} 
\includegraphics{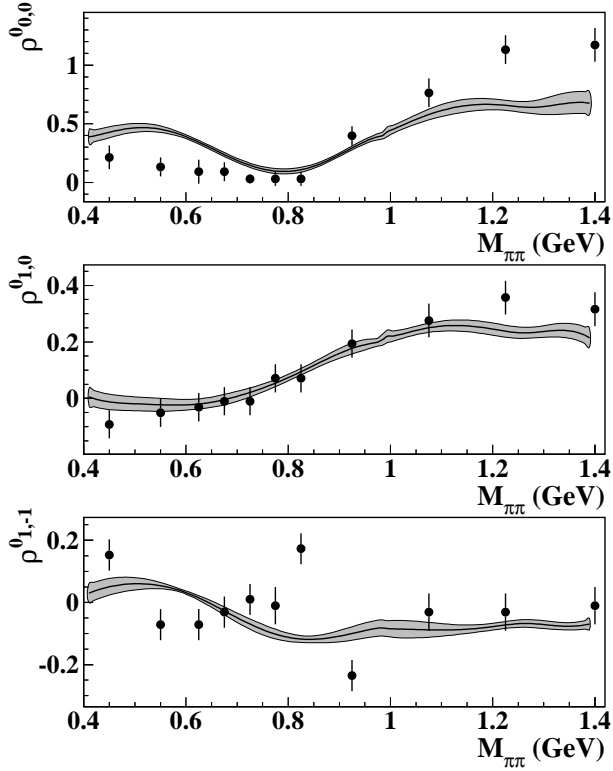}
\caption[]{
Spin density matrix elements for the $P$-wave in the  $3.0 <E_\gamma< 3.2$~GeV  and $0.4<-t<0.5$~GeV$^2$ bin.
The black dots are data points from Ref.~\cite{Ballam_1}, taken in a similar kinematic bin 
($E_\gamma \sim 2.8$~GeV and $0.02<-t<0.4$~GeV$^2$).
}
\label{fig:rho_super}
\end{figure}

\subsection{The spin density matrix elements} 
\begin{figure} 
\vspace{10.cm} 
\includegraphics{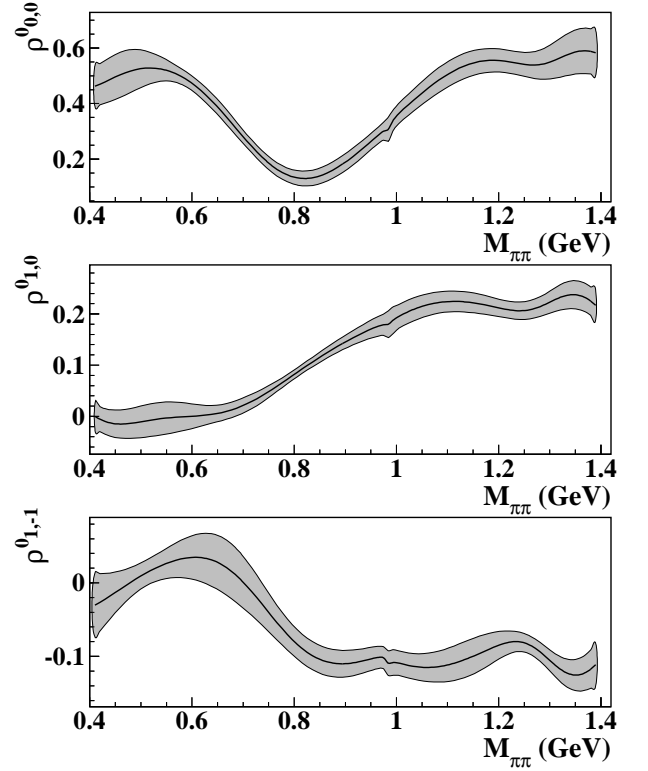}
\caption[]{Spin density matrix elements for the $P$-wave in the  $3.2 <E_\gamma< 3.4$~GeV  and $0.5<-t<0.6$~GeV$^2$ bin.}
\label{fig:rho_decay_p}
\end{figure}

\begin{figure} 
\vspace{7.5cm} 
\includegraphics{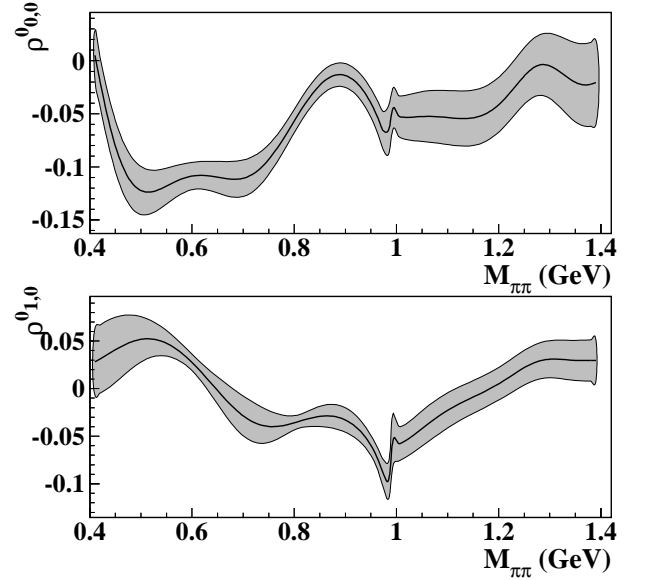}
\caption[]{Spin density matrix elements for the interference between $S$- and $P$-waves in the  $3.2 <E_\gamma< 3.4$~GeV  and $0.5<-t<0.6$~GeV$^2$ bin.}
\label{fig:rho_decay_ps}
\end{figure}

From the production amplitudes derived by the fit, we calculated the spin density matrix elements~\cite{Schilling:1969um} for the 
$P$-wave and the interference between the $S$- and $P$-waves.
Some selected results are shown in Figs.~ \ref{fig:rho_super},~\ref{fig:rho_decay_p} and ~\ref{fig:rho_decay_ps}.
Since these observables do not depend on the photon flux normalization, the error bands do not include 
the 10\% uncertainty mentioned above.
The whole set of spin density matrix elements resulting from this analysis  is  available at the Jefferson Lab~\cite{jlab-db} and the Durham~\cite{dhuram-db} databases.

Comparisons of our measurements at $3.0 <E_\gamma< 3.2$~GeV  and $0.4<-t<0.5$~GeV$^2$ 
with existing data from Refs.~\cite{Ballam_1,Ballam_2}  in a similar kinematic domain 
($E_\gamma \sim 2.8$ GeV and $0.02<-t<0.4$~GeV$^2$) are shown in Fig.~ \ref{fig:rho_super}.
As expected, the two matrix elements $\rho_{10}$ and  $\rho_{11}$ agree very well since they have a weak
dependence on $-t$, while $\rho_{00}$ shows a similar behavior, but with different values as it is  more sensitive 
to the momentum transfer. If one compares 
the larger $-t$ bins we measured, the differences increase, showing that extrapolating 
our data to lower $-t$  would probably give  good agreement with 
previous measurements.

As shown in Fig.~\ref{fig:rho_decay_ps}, around $M_{\pi \pi}=980$~MeV an interference pattern clearly  
shows up in the $S$-$P$ wave interference term, corresponding to the contribution from the 
$f_0(980)$ meson.

\subsection{Differential cross sections} 
\begin{figure}[ht]
\vspace{6.cm} 
\includegraphics{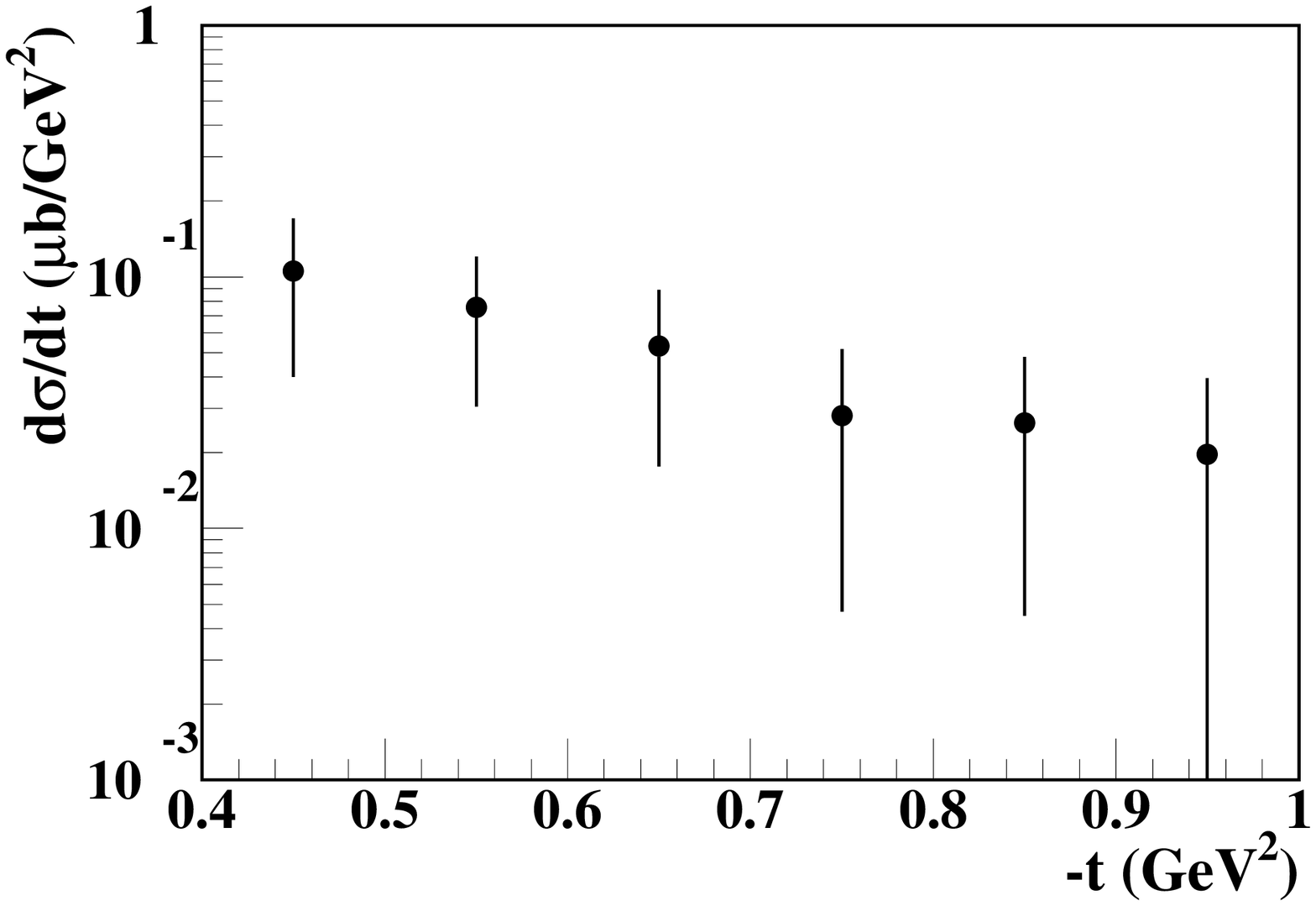}
\caption[]{Differential cross section  $d\sigma/dt$   for the  $S$-wave  in the $M_{\pi\pi}$ 
range $0.98\pm0.04$~GeV and photon energy range $E_\gamma=3.0 - 3.8$ GeV.} 
\label{fig:dsdt-s-f0-no-fsi}
\end{figure}

\begin{figure}[ht]
\vspace{6.cm} 
\includegraphics{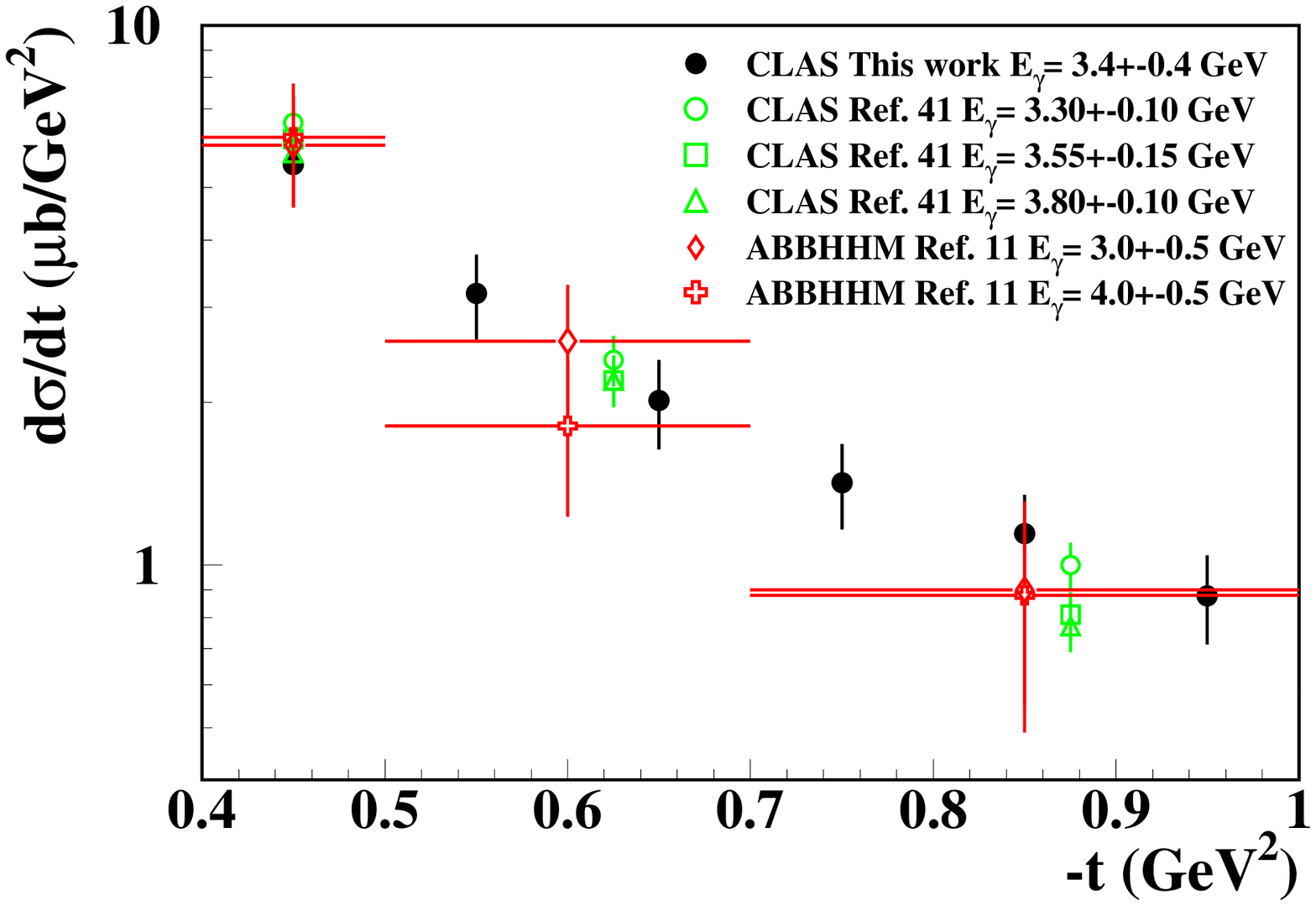}
\caption[]{ Differential cross section  $d\sigma/dt$   for the  $P$-wave  in the $M_{\pi\pi}$ 
range 0.4-1.2  GeV and photon energy range $E_\gamma=3.0 - 3.8$ GeV.} 
\label{fig:dsdt-p-rho-no-fsi}
\end{figure}

\begin{figure}[ht]
\vspace{6.cm} 
\includegraphics{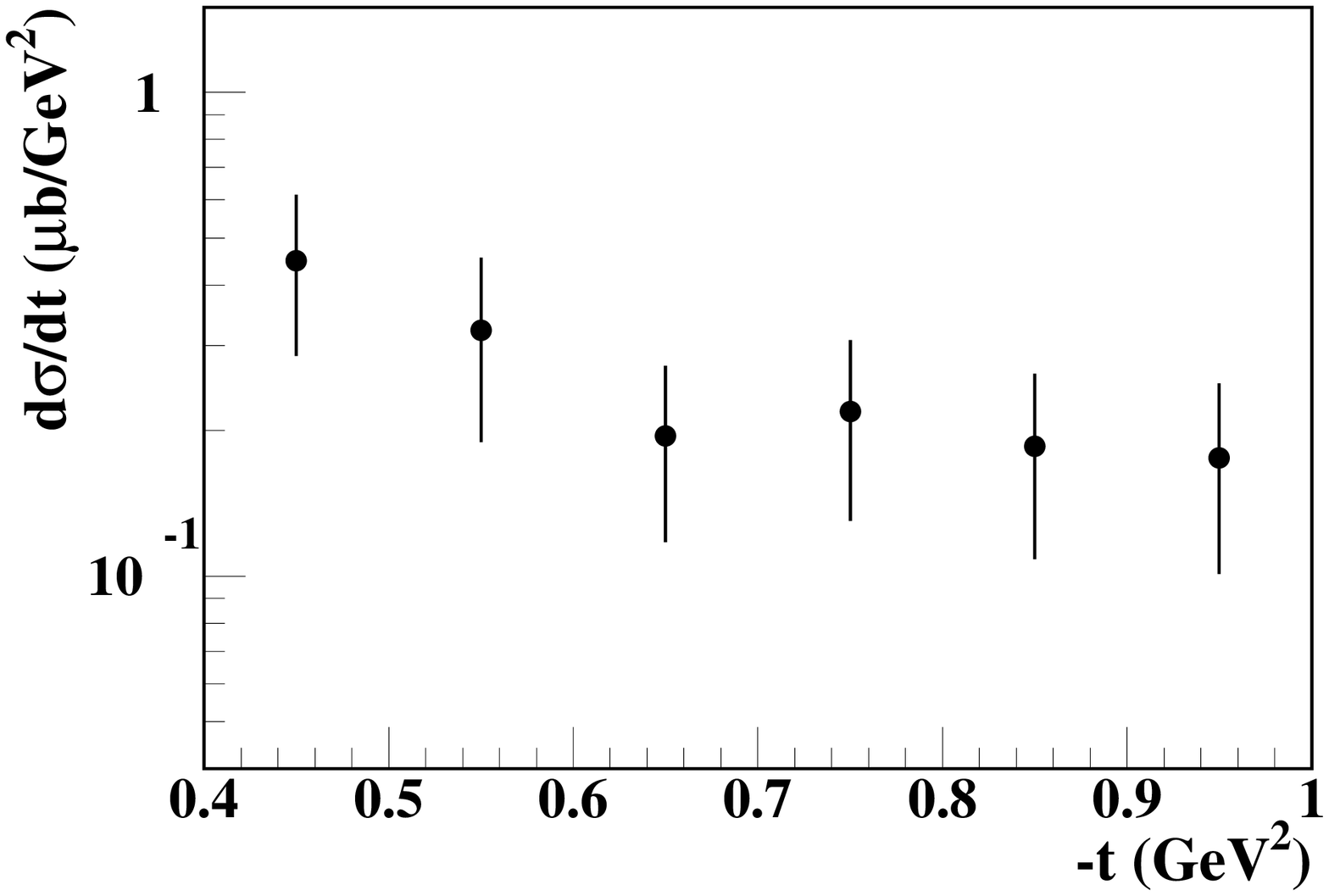}
\caption[]{Differential cross section  $d\sigma/dt$   for the  $D$-wave  in the $M_{\pi\pi}$
range $1.275\pm0.185$  GeV and photon energy range $E_\gamma=3.0 - 3.8$ GeV. } 
\label{fig:dsdt-d-f2-no-fsi}
\end{figure}

The differential cross sections $[d\sigma/dt]_{l-wave}$ for individual waves and mass resonance  regions 
were obtained integrating the  corresponding amplitudes. 
The  cross sections in the mass regions of the $f_0(980)$, $\rho$, and $f_2(1270)$ mesons were obtained
integrating the $S$-, $P$- and $D$-waves in the mass ranges $0.98\pm0.04$~GeV, 0.4-1.2 GeV, and $1.275\pm0.185$~GeV,
respectively. These are shown in Figs.~\ref{fig:dsdt-s-f0-no-fsi},~\ref{fig:dsdt-p-rho-no-fsi} and ~\ref{fig:dsdt-d-f2-no-fsi}
in the photon energy range 3.0-3.8~GeV. As mentioned previously, the $P$-wave is completely dominated by the $\rho$ meson production,
and therefore the integrated cross section can be directly compared to the world's data for the $\gamma p \to p \rho$ 
reaction~\cite{Battaglieri,ABBHHM}. It should be noticed that the previous cross sections were evaluated without performing a partial wave analysis 
but fitting the mass-dependent cross section with a relativistic Breit-Wigner plus a smooth polynomial function to separate the 
resonance from the background. The good agreement shown in Fig.~\ref{fig:dsdt-p-rho-no-fsi} gives confidence in the partial wave 
analysis. As expected, the  $S$-wave photoproduction is suppressed compared to  the $P$-wave by more than an order of magnitude,
reflecting the different mechanisms that lead  to scalar and vector meson photoproduction: in Regge theory the latter is dominated by
 Pomeron exchange, while the former is dominated by the  exchange of  reggeons
that become suppressed as the energy increases.

\section{\label{sec:sum} Summary}
In summary, we have performed a partial wave analysis of the reaction  $\gamma p \to p \pi^+ \pi^-$ in the photon energy range 3.0-3.8~GeV
and momentum transfer range $-t=0.4-1.0$~GeV$^2$. Moments of the di-pion angular distribution, defined as bi-linear functions of partial wave amplitudes,
were fitted to the experimental data with an unbinned likelihood procedure. Different parametrization bases were used and detailed systematic checks 
were performed to insure the reliability of the analysis procedure. We extracted moments $\langle Y_{LM}\rangle$ with $L \le 4$ and $M \le 2$ using amplitudes with $l \le 3$ (up to $F$-waves). 
Using a dispersion relation, unitarity constraint, and 
phase shifts and inelasticities of $\pi\pi$ scattering, the production amplitudes were expressed in a simplified form,
where the unknown part was expanded  in a Taylor series. The coefficients were fitted to the experimental moments to extract the
$S$-, $P$-,
$D$-, and $F$-waves in the $M_{\pi\pi}$ range 0.4-1.4 GeV.\\

The moment $\langle Y_{00} \rangle$ is dominated by the $\rho(770)$ meson contribution in the $P$-wave,
while the moments $\langle Y_{10} \rangle$ and $\langle Y_{11} \rangle$  show  contributions of the $S$-wave  through 
interference with the $P$-wave. The clear structure at $M_{\pi\pi} \sim 1\mbox{ GeV}$ seen in such experimental 
moments and in the $S$-wave amplitude
is evidence of a resonance contribution that we interpret as the $f_0(980)$. This is the first observation of the 
$f_0(980)$ scalar meson in photoproduction. A contribution from the $f_2(1270)$ tensor meson was observed in the $D$-wave, 
while no resonant structures
were seen in the $F$-wave. The cross sections of individual partial waves in the mass range of the $\rho(770)$,  $f_0(980)$, and $f_2(1270)$ were 
computed. Finally, the spin density matrix elements for the $P$-wave  were evaluated, finding good agreement with previous measurements, and 
for the first time, the  $S-P$ interference term was extracted.

\section{\label{sec:ack} Acknowledgments}
We would like to acknowledge the outstanding efforts of the staff of the Accelerator
and the Physics Divisions at Jefferson Lab that made this experiment possible. 
This work was supported in part by  the  Italian Istituto Nazionale di Fisica Nucleare, 
the French Centre National de la Recherche Scientifique
and Commissariat \`a l'Energie Atomique, the UK Science and Technology Facilities Research Council (STFC),
the U.S. Department of Energy and National Science Foundation, 
and the Korea Science and Engineering Foundation.
The Southeastern Universities Research Association (SURA) operates the
Thomas Jefferson National Accelerator Facility for the United States
Department of Energy under contract DE-AC05-84ER40150.
\appendix
\section{}\label{app:A}
The explicit expressions for the moments, defined in  Eq.~\ref{eq:mom}  in terms of partial waves, given Eq.~\ref{partial}, truncated to the $L=3$ ($F$) wave are given by, 

\begin{widetext}
\begin{eqnarray*}
\langle Y_{00} \rangle &=&  |S|^2 + |P_-|^2 + |P_0|^2 + |P_+|^2 + |D_-|^2 + |D_0|^2 + |D_+|^2 + |F_-|^2 + |F_0|^2 + |F_+|^2 \\
\langle Y_{10} \rangle &=&  S P_0^* + P_0 S^* + \sqrt{3\over 5} \left(P_- D_-^*  + P_-^* D_-  +  P_+ D_+^*  + D_+ P_+^*\right)
                          + \sqrt{4\over 5} \left( P_0 D_0^* + D_0 P_0^* \right) \\
                      &+&  \sqrt{ {24} \over {35}} \left( D_- F_-^* + F_- D_-^*  + D_+ F_+^* + F_+ D_+^* \right) 
                          + \sqrt{ {216} \over {280}} \left( D_0 F_0^* + F_0 D_0^* \right)  \\
\langle Y_{11} \rangle &=&   \left( -P_- S^* - S P_-^* + P_+ S^* + S P_+^* \right) 
                          + \sqrt{1\over {20}} \left( P_- D_0^* + D_0 P_-^*  - P_+ D_0^*  - D_0 P_+^* \right)\\ 
                      &+&  \sqrt{3\over {20}} \left( -P_0 D_-^* - D_- P_0^* + P_0 D_+^* + D_+ P_0^* \right)
                          +\sqrt{9\over {140}} \left( D_- F_0^* + F_0 D_-^* - D_+ F_0^* - F_0 D_+^* \right)\\ 
                      &+&  \sqrt{ 9\over {70}} \left( -D_0 F_-^* - F_- D_0^* +  D_0 F_+^*  + F_+ D_0^* \right) \\
\langle Y_{20} \rangle &=&   S D_0^* + D_0 S^* + \sqrt{1\over 5} \left( 2 |P_0|^2 -|P_-|^2   - |P_+|^2 + |F_-|^2 + |F_+|^2 \right) 
                          + \sqrt{{18}\over {35}} \left( P_- F_-^* + F_- P_-^* +  P_+ F_+^* + F_+ P_+^* \right)\\
                      &+&  \sqrt{ {27}\over {35}} \left( P_0 F_0^* + F_0 P_0^* \right)  + \sqrt{5\over {49}} \left( |D_-|^2 + |D_+|^2 \right) 
                          + \sqrt{{20}\over {49}} |D_0|^2 + \sqrt{{16}\over {45}} |F_0|^2 \\
\langle Y_{21} \rangle &=& {1\over 2} \left( S D_+^* + D_+ S^* - S D_-^* - D_- S^* \right) 
                          + \sqrt{ 3\over {20}} \left( P_0 P_+^* + P_+ P_0^*  - P_- P_0^* - P_0 P_-^* \right)\\ 
                      &+&  \sqrt{9 \over {140}} \left( P_- F_0^* + F_0 P_-^* - P_+ F_0^*- F_0 P_+^* \right)
                          + \sqrt{6\over {35}} \left( P_0 F_+^* + F_+ P_0^* - P_0 F_-^* - F_- P_0^* \right)\\ 
                      &+&  \sqrt{5 \over {196}} \left(D_0 D_+^* + D_+ D_0^*  - D_0 D_-^*  - D_- D_0^*  \right) 
                          + \sqrt{1\over {90}} \left( F_0 F_+^* + F_+ F_0^* - F_0 F_-^* - F_- F_0^* \right)  \\
\langle Y_{22} \rangle &=&  \sqrt{3 \over {10}} \left( P_- P_+^* + P_+ P_-^* \right) 
                          + \sqrt{3 \over {140}} \left( P_- F_+^* + F_+ P_-^*  + P_+ F_-^* + F_- P_+^* \right)  \\
                      &+&  \sqrt{ 4 \over {30}} \left( - F_+ F_-^* - F_- F_+^* \right) + \sqrt{3 \over {196}} \left( - D_- D_+^* - D_+ D_-^* \right) \\
\langle Y_{30} \rangle &=&  S F_0^* + F_0 S^* + \sqrt{ {18}\over {70}} \left( - P_- D_-^* - D_- P_-^* 
                          - P_+ D_+^*  - D_+ P_+^* \right) + \sqrt{{108} \over {140}} \left( P_0 D_0^* + D_0 P_0^* \right) \\
                      &+&  \sqrt{ 2\over {45}} \left( D_- F_-^* + F_- D_-^*  + D_+ F_+^* + F_+ D_+^* \right) 
                          + \sqrt{ {16} \over {45}} \left( D_0 F_0^* + F_0 D_0^* \right)  \\
\langle Y_{31} \rangle &=&  {1\over 2} \left( S F_+^* + F_+ S^* - S F_-^* - F_- S^* \right) 
                          + \sqrt{{18} \over {140}} \left( P_+ D_0^*  + D_0 P_+^* - P_- D_0^* - D_0 P_-^* \right)\\ 
                      &+& \sqrt{ 6 \over {35}} \left( P_0 D_+^* + D_+ P_0^* - P_0 D_-^* - D_- P_0^* \right) 
                          + \sqrt{ 1\over {90}} \left( D_+ F_0^* + F_0 D_+^* - D_- F_0^*  - F_0 D_-^* \right)\\ 
                      &+& \sqrt{ 1\over {20}} \left( D_0 F_+^* + F_+ D_0^* -D_0 F_-^*  - F_- D_0^* \right)  \\
\langle Y_{32} \rangle &=&   \sqrt{3 \over {14}} \left( -P_+ D_-^* - D_- P_+^*  -P_- D_+^* - D_+ P_-^* \right) 
                          + \sqrt{1 \over {12}} \left( - D_+ F_-^* - F_- D_+^* - D_- F_+^* - F_+ D_-^* \right)  \\
\langle Y_{40} \rangle &=&   \sqrt{2\over 7} \left( - P_+ F_+^* - F_+ P_+^*  - P_- F_-^* - F_- P_-^* \right) 
                          + \sqrt{{16}\over {21}} \left( P_0 F_0^* + F_0 P_0^* \right) 
                          + \sqrt{{16} \over {49}}  \left( - |D_+|^2 - |D_-|^2 \right) \\
                      &+&  \sqrt{{36}\over {49}} |D_0|^2  + \sqrt{{36} \over {121}} |F_0|^2 + \sqrt{1\over {121}} \left( |F_+|^2 + |F_-|^2 \right) \\
\langle Y_{41} \rangle &=&  \sqrt{5\over {42}} \left( P_+ F_0^* + F_0 P_+^* - P_- F_0^*  - F_0 P_-^* \right) 
                          + \sqrt{5\over {28}} \left( P_0 F_+^* + F_+ P_0^*  - P_0 F_-^*  - F_- P_0^* \right)   \\
                      &+&.\sqrt{{ 30}\over {196}} \left( D_0 D_+^* + D_+ D_0^*  - D_- D_0^* - D_0 D_-^*\right) 
                          + \sqrt{{30} \over {968}} \left( F_0 F_+^* + F_+ F_0^* - F_0 F_-^* - F_- F_0^* \right) \\
\langle Y_{42} \rangle &=&   \sqrt{5\over {28}} \left( - P_+ F_-^* - F_- P_+^* - P_- F_+^*  - F_+ P_-^* \right)  
                          + \sqrt{{10} \over {49}} \left( -D_- D_+^* - D_+ D_-^* \right) \\
                      &+&\sqrt{ {10} \over {121}} \left( - F_- F_+^* - F_+ F_-^* \right)  \\
\end{eqnarray*}
\end{widetext}

It follows from Eq.~\ref{eq:mom} that the $\langle Y_{00} \rangle$ moment is normalized by the differential cross section via, 
\begin{equation} 
 \langle Y_{00} \rangle  =  \int d\Omega_\pi \frac{d\sigma}{dt dM_{\pi\pi} d\Omega_\pi}. 
\end{equation}

\end{document}